\documentclass[fleqn,usenatbib]{mnras}

\usepackage{newtxtext,newtxmath}

\usepackage[T1]{fontenc}
\usepackage{ae,aecompl}

\usepackage{graphicx}	
\usepackage{amsmath}	
\usepackage{amssymb}	
\usepackage{subfigure}  

\newcommand{\appropto}{\mathrel{\vcenter{
  \offinterlineskip\halign{\hfil$##$\cr
    \propto\cr\noalign{\kern2pt}\sim\cr\noalign{\kern-2pt}}}}}
 \usepackage[dvipsnames]{xcolor}

\title[\emph{Gaia} DR2 in Action Space]{The Galactic Disc in Action Space as seen by \emph{Gaia} DR2}

\author[W. H. Trick et al.]{
Wilma H. Trick,$^{1}$\thanks{E-mail: trick@mpa-garching.mpg.de}
Johanna Coronado,$^{2}$
and Hans-Walter Rix$^{2}$
\\
$^{1}$Max-Planck-Insitut f\"ur Astrophysik, Karl-Schwarzschild-Str. 1, D-85748 Garching b. M\"unchen, Germany\\
$^{2}$Max-Planck-Insitut f\"ur Astronomie, K\"onigstuhl 17, D-69117 Heidelberg, Germany
}

\date{Accepted 2019 January 16. Revised 2019 January 16. Received 2018 May 10}

\pubyear{2015}

\begin{document}
\label{firstpage}
\pagerange{\pageref{firstpage}--\pageref{lastpage}}
\maketitle

\begin{abstract}
The quality and quantity of 6D stellar position-velocity measurements in the second \emph{Gaia} data release's radial velocity sample (DR2/RVS) allow us to study small-scale structure in the orbit distribution of the Galactic disc beyond the immediate Solar neighbourhood. We investigate the distribution of orbital actions $(J_R,L_z,J_z)$ of $\sim 3.9$ million stars within $1.5~\text{kpc}$ of the Sun, for which precise actions can be calculated from \emph{Gaia} DR2 alone. In the $n(J_R,L_z)$ distribution, we identify at least seven stellar overdensities at high $J_R$ that lie along lines of constant slope $\Delta J_R / \Delta L_z$ and appear to be present in the stellar distribution out to $\sim1.5~\text{kpc}$ at the same location in $(L_z,J_R)$. The known moving groups in the $(U,V)$-plane of the Solar neighbourhood are local manifestations of this extended system of orbit substructure at low $J_R$, as expected. The $n(J_R,L_z)$ features are most prominent among stars within $|z|\lesssim 500~\text{pc}$ from the Galactic plane and with lower than average $J_z$. These features frequently coincide with $(L_z,J_R)$ regions of dramatic imbalance between stars moving in and out, suggesting that stars are not phase-mixed along orbits or on resonant orbits. Some of these $n(J_R,L_z)$ ridges resemble features expected from rapid orbit diffusion along particular $(J_R,L_z)$-directions in the presence of various resonances. Orbital action and angle space of stars in \emph{Gaia} DR2 is therefore highly structured over kpc scales, and appears to be very informative for modelling studies of non-axisymmetric structure and resonances in the Galactic disc.
\end{abstract}




\section{Introduction}

The orbital actions $(J_R,L_z,J_z)$ are a powerful tool to characterize the orbits of stars in the Galaxy's gravitational potential (\citealt{2008gady.book.....B}, \S3.5). In axisymmetric potentials, they are integrals of motion and quantify the amount of oscillation of the star along its (short-axis tube) orbit in Galactocentric $(R,\phi,z)$ direction, respectively. This makes them ideal orbit labels and ideal parameters for distribution functions (DFs) that describe stellar components of the Galaxy, e.g. the disc \citep{2011MNRAS.413.1889B,2015MNRAS.449.3479S} and the halo \citep{2015MNRAS.447.3060P,2016MNRAS.460.1725D}, as well as dark matter particles in the halo \citep{2015MNRAS.454.3653B}. Recently, several studies have made use of action-based dynamical modelling to learn more about the Milky Way's gravitational potential \citep{2013ApJ...779..115B,2014MNRAS.445.3133P,2016ApJ...830...97T} and chemo-orbital structure \citep{2013ApJ...779..115B,2015MNRAS.449.3479S,2016MNRAS.463.3169D}. These studies modelled the Galaxy as axisymmetric and phase-mixed, i.e., stars are evenly distributed along orbits (in orbital angle space). But orbital actions are also powerful diagnostics of non-axisymmetric perturbations \citep{1988MNRAS.230..597B,2012ApJ...751...44S,2015MNRAS.449.1982F,2016MNRAS.457.2569M,2017MNRAS.465.1443M,2017MNRAS.466L.113M,2017MNRAS.471.4314M,2018MNRAS.474.2706B}. Both long-lived and stochastic perturbations can lead to orbit diffusion that forms distinct features in orbit space. In particular, ridges in $(L_z,J_{R})$ space develop, seen both in numerical experiments (e.g., \citealt{2012ApJ...751...44S}) and analytic (Fokker-Planck) models (e.g., \citealt{2015ApJ...806..117F}). Very locally, this manifests itself in streams and structure in $(U,V,W)$ velocity space. In particular, many of the moving groups observed in the $(U,V)$-plane of the Solar neighbourhood (\citealt{1998AJ....115.2384D}; \citealt{1996AJ....112.1595E}, and references therein)---among them the Hyades and the Hercules stream---are expected to have such a dynamic origin (e.g., \citealt{2000AJ....119..800D,2005A&A...430..165F,2007A&A...461..957F,2008A&A...483..453F}; see Appendix \ref{sec:reviewing_moving_groups} for more details).

The second \emph{Gaia} data release (DR2) on April 25, 2018 \citep{2016A&A...595A...1G,2018A&A...616A...1G} provided consistent high-precision astrometric measurements of positions and proper motions, as well as measurements of radial velocities and stellar parameters by the Radial Velocity Spectrometer (RVS) \citep{2018arXiv180409372K} for millions of stars, ushering in a new era in Galactic astronomy. Arches and shells have been found in the disc's \emph{Gaia} DR2 velocity space \mbox{\citep{2018A&A...616A..11G,2018Natur.561..360A}}, `snail shells' and ridges in space-velocity diagrams \citep{2018Natur.561..360A,2018MNRAS.479L.108K}, some of them indicating incomplete vertical phase-mixing \citep{2018Natur.561..360A,2018MNRAS.481.1501B,2019MNRAS.tmp..222B,2018RNAAS...2b..32M}. While much substructure in the Galactic disc was indeed expected (cf. discussion in \citealt{2009ApJ...700.1794B}), observed in many previous observations (e.g., \citealt{2018MNRAS.478.3809S,2018arXiv180407530K}), and is also predicted from cosmological galaxy simulations \citep{1999Natur.402...53H}, we have never seen it in such unprecedented detail as with \emph{Gaia} DR2.

In this work, we set out to qualitatively investigate the orbital distribution of stars within $\sim 1.5$~kpc from the Sun, by estimating the actions of all \emph{Gaia} DR2/RVS stars that provide full 6D phase-space measurements. Our goal is to chart new and known substructure in {\it orbit space} and to get a feeling for (a) the differences in stellar occupancy on the different orbits and (b) the extent of non-axisymmetry in these substructures. Both will be very important for future modelling attempts of the Milky Way disc (e.g., \citealt{2017ApJ...839...61T}).

Mapping substructure in orbit space, rather than $(X,Y,Z;U,V,W)$ configuration space, has several advantages: it maps the 6D phase-space information of data into a canonical coordinate system that has straightforward interpretation; and in going beyond tiny volumes (say, $d<200~\text{pc}$) in the Milky Way, it is arguably the best way to identify stars on the same orbits. 

Analogous work on substructure in action space has already been performed by \citet{2010MNRAS.409..145S}, albeit with data from the Hipparcos mission \citep{2000A&A...355L..27H} and the Geneva-Copenhagen survey \citep{2009A&A...501..941H} that was both lacking in quality and quantity as compared to \emph{Gaia} DR2. \citet{2010MNRAS.409..145S} found that the Hyades stream maps into an overdensity in action space that could be caused by resonances of the Galactic bar and/or spiral arms. This work was later on extended by \citet{2011MNRAS.418.1565M}. To account for selection effects in angle space, he compared the data to a phase-mixed disc model. He also showed that Sirius and the Pleiades correspond to overdensities in action space.

This paper is organized as follows. In Section \ref{sec:sample_selection}, we present the stellar sample that we are going to investigate. In Section \ref{sec:actions}, we give a short introduction to actions and to action estimation---which the experienced reader is encouraged to skip---and mention the assumed gravitational potential model and estimation method employed in this work. To illustrate selection effects as well as our naive expectations for the action distribution in a perfectly axisymmetric Galaxy, we generate and discuss mock data in Section \ref{sec:mock_data}. In Section \ref{sec:extended_orbit_structure}, we present the distribution of \emph{Gaia} DR2/RVS stars in action space in different spatial bins  and argue that this reveals the large-scale orbit structure of the Galactic disc beyond the Solar neighbourhood. In Section \ref{sec:moving_groups}, we compare the extended orbit structure in action space to the known moving groups in velocity space of the Solar neighbourhood. Sections \ref{sec:asymmetric_vR} and \ref{sec:mean_Jz} show how the extended orbit substructure is related to asymmetries in the $v_R$ distribution and the average vertical action of the stars. We discuss our results and conclude in Section \ref{sec:discussion_conclusion}.

\section{Data and Method}

The second data release of the \emph{Gaia} satellite \citep{2016A&A...595A...1G,2018A&A...616A...1G} provides full 6D space coordinates for $N_\text{RVS}=$ 7,224,631 stars: 2D positions (RA,Dec), parallaxes $\varpi$, proper motions
$(\mu_{\mathrm{RA*}},\mu_{\mathrm{Dec}})$ \citep{2018A&A...616A...2L}, 
and radial line-of-sight velocities $v_\text{los}$ down to $G=13$ mag measured by the Radial Velocity Spectrometer (RVS) instrument \citep{2018arXiv180409372K}. The sample size is essentially set by the availability of radial velocity estimates. This data set is by far the largest, consistent sample to estimate precise orbits (i.e. orbital actions) for stars in an extended region around the Sun.

\subsection{6D stellar phase-space data from \emph{Gaia} DR2/RVS}

\subsubsection{Sample selection} \label{sec:sample_selection}

Meaningful action estimates require high-precision data, in particular good distance measurements \citep{2018MNRAS.481.2970C}. We restrict our analysis therefore to the 3,872,301 stars within 1.5~\text{kpc} (with positive parallaxes) of which the vast majority have relative parallax uncertainties $\delta\varpi/\varpi < 0.05$ (see Figure \ref{fig:distance_bins}), which roughly translates into a relative distance error of $5\%$. This is precise enough for us to use $d=1/\varpi$ as an acceptable distance estimate (however, see \citealt{2018AJ....156...58B,2018A&A...616A...9L}).

We do not correct the parallaxes by the zero-point offset (e.g. \citealt{2018A&A...616A...9L}), but the error introduced by this is mostly less than the parallax measurement uncertainty for this local sample. We also do not impose any additional quality cuts on the data. In Appendix \ref{sec:data_quality}, we further motivate this choice and our distance cut at $1.5~\text{kpc}$.

\begin{figure*}
\centering
\subfigure[$\sim370,000$ stars in the Solar neighbourhood within $1/\varpi < 200~\text{pc}$. \label{fig:sol_neighbourhood_200pc}]{\includegraphics[width=0.9\textwidth]{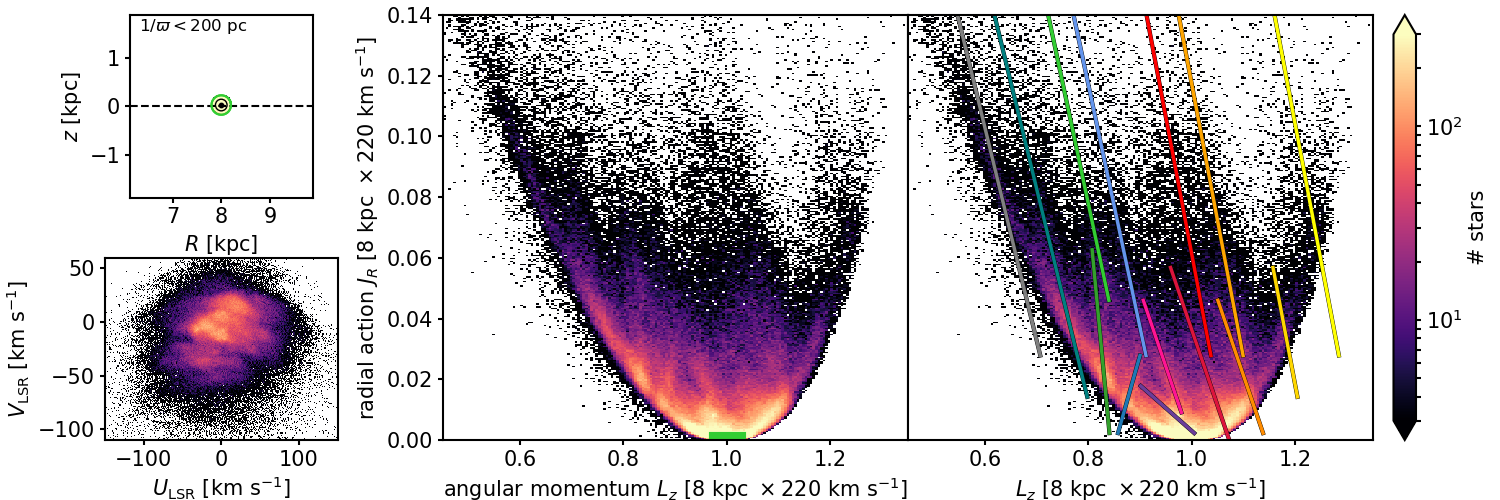}} %
\subfigure[$\sim1.7$ Million stars from \emph{Gaia} DR2/RVS with $200~\text{pc} < 1/\varpi < 600~\text{pc}$. \label{fig:sol_neighbourhood_600pc}]{\includegraphics[width=0.9\textwidth]{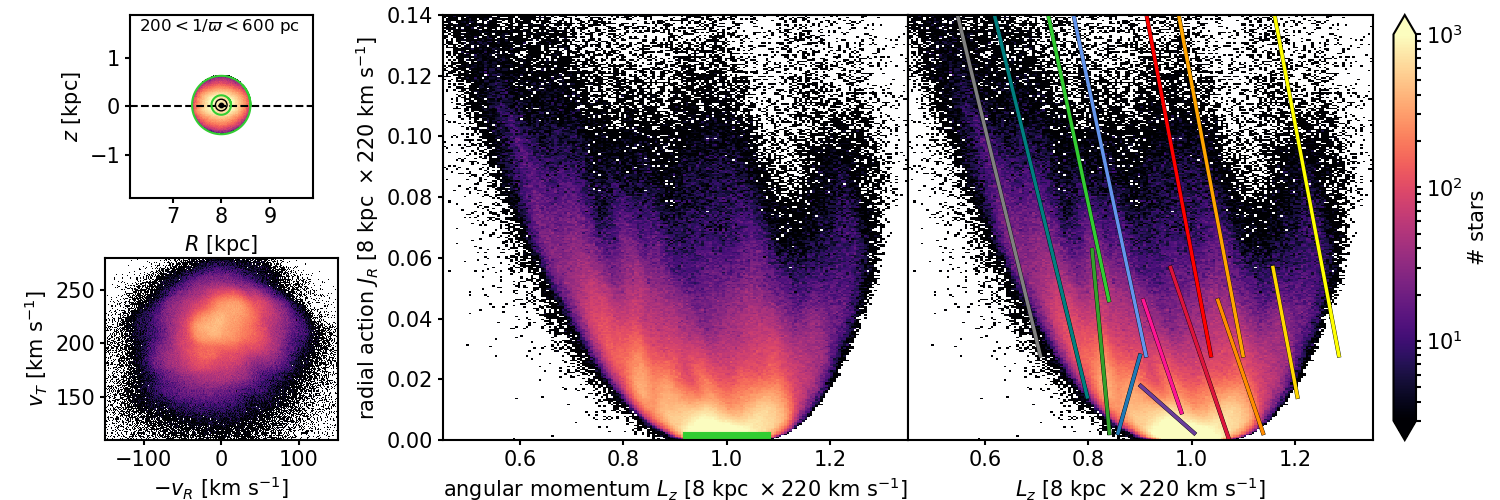}} %
\subfigure[$\sim1.7$ Million stars from \emph{Gaia} DR2/RVS with $600~\text{pc} < 1/\varpi < 1.5~\text{kpc}$. \label{fig:sol_neighbourhood_1500pc}]{\includegraphics[width=0.9\textwidth]{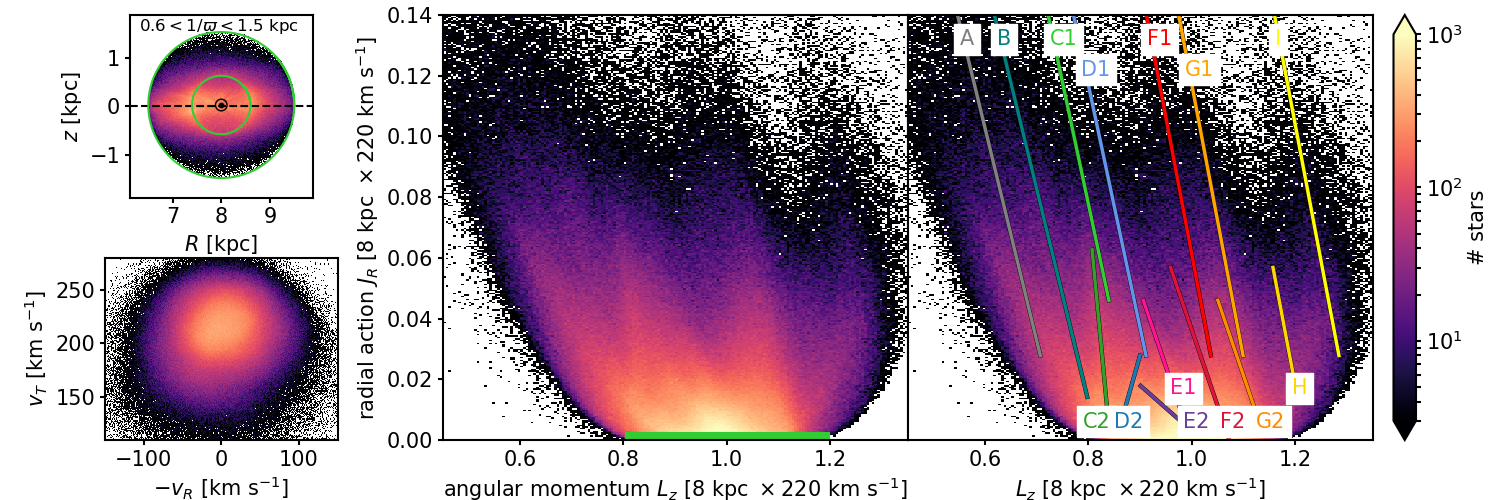}}
\caption{The action distribution $n(L_z,J_R)$ of the \emph{Gaia} DR2/RVS sample described in the text, overplotted with the locations in $(L_z,J_R)$ that we found by eye to correspond to the overdensities of the extended orbit substructure in the Galactic disc (coloured lines in the right-hand panels; the labels in Panel \ref{fig:sol_neighbourhood_1500pc} refer to Table \ref{tab:extended_orbit_substructure}). We show this independently for the three different radial shells around the Sun ($1/\varpi < 200~\text{pc}$, Figure \ref{fig:sol_neighbourhood_200pc}; $200~\text{pc} < 1/\varpi  \lesssim 600~\text{pc}$, Figure \ref{fig:sol_neighbourhood_600pc}; $600~\text{pc} < 1/\varpi \lesssim 1.5~\text{kpc}$, Figure \ref{fig:sol_neighbourhood_1500pc}). In addition, we show the corresponding distributions in the Galactocentric meridional plane (left top), and radial and azimuthal velocities in the plane of the disc (left bottom; $(U,V)$ with respect to the LSR for the Solar neighbourhood in Panel \ref{fig:sol_neighbourhood_200pc}, and the equivalent Galactocentric velocities $-v_{R}$ and $v_{T}$ in Panels \ref{fig:sol_neighbourhood_600pc}-\ref{fig:sol_neighbourhood_1500pc}). The bin sizes in the 2D histograms are: $15~\text{pc}$ and $1~\text{km s}^{-1}$ in position and velocity, respectively, $0.005L_{z,0}$ in $L_z$, and $0.0005L_{z,0}$ in $J_R$. In the upper left panels, the location of the Sun is marked by an $\odot$, the Galactic plane by a dashed line, and annuli in Solar distance $1/\varpi$ in green. The green, horizontal bar in the middle panels at $J_R\sim0$ represents the radial range of the samples in $L_z$ (see Section \ref{sec:survey_volume} for details). This figure illustrates (i) the richness of orbit substructure seen in \emph{Gaia} DR2 and (ii) that it appears to be part of the same extended orbit structure showing up at the same $(L_z,J_R)$ and with the same slopes (summarized in Table \ref{tab:extended_orbit_substructure}) independent of the spatial bin considered.}
\label{fig:sol_neighbourhood}
\end{figure*}

\subsubsection{Galactocentric coordinates}

We first convert the 6D coordinates from observables into $(X,Y,Z)$ positions and the corresponding heliocentric $(U_\text{HC},V_\text{HC},W_\text{HC})$ velocities, centred at the position and velocity of the Sun, using the coordinate transformations in \citet{2015ApJS..216...29B}'s \texttt{galpy} package. $U$ moves towards the Galactic centre, $V$ is the velocity component in the direction of Galactic rotation, and $W$ towards the Galactic North pole. For the Sun's motion with respect to the Local Standard of Rest (LSR), we use $(U_\odot,V_\odot,W_\odot) = (11.1, 12.24, 7.25)~\text{km s}^{-1}$ \citep{2010MNRAS.403.1829S}, and convert the heliocentric $(U_\text{HC},V_\text{HC})$ to $(U_\text{LSR},V_\text{LSR})$.

Following \citet{2015ApJS..216...29B}, we assume $R_\odot = 8~\text{kpc}$ and $v_\text{circ}(R_\odot) = 220~\text{km s}^{-1}$ \citep{2012ApJ...759..131B} for the Sun's distance to the Galactic centre and the circular velocity of the Milky Way at the Solar radius. The height above the Galactic plane of the Sun is assumed to be $z_\odot = 25~\text{pc}$ \citep{2008ApJ...673..864J}. Using this, we can transform the $(X,Y,Z,U_\text{LSR},V_\text{LSR},W_\text{LSR})$ into Galactocentric cylindrical $(R,\phi,z,v_R,v_T,v_z)$ coordinates (assuming the LSR moves on a circular orbit).

The distributions of stars in the Galactocentric meridional plane, $n(R,z)$, and the in-plane velocities, $n(U,V)$ or $n(-v_R,v_T)$,\footnote{The $(U_\text{LSR},V_\text{LSR})$ and Galactocentric $(v_R,v_T)$ coordinates differ by the sign in the definition of $U_\text{LSR}$ and $v_{R}$, and by the shift of $v_\text{circ}(R) \sim 220~\text{km s}^{-1}$ between $v_\text{LSR}$ and $v_T$.} are shown for different distance bins from the Sun in the upper left panels of Figure \ref{fig:sol_neighbourhood}. The stellar sample with distances $1/\varpi < 200~\text{pc}$ from the Sun (Figure \ref{fig:sol_neighbourhood_200pc}) covers the spatial extent of the Hipparcos mission. Outside of $\sim1.5~\text{kpc}$, the data set becomes considerable incomplete due to the magnitude limit of the RVS instrument (\citealt{2018arXiv180409372K}; see also Appendix \ref{sec:data_quality}). The $(U_\text{LSR},V_\text{LSR})$ plane of the local neighbourhood reveals the well-known signatures of the moving groups in unprecedented detail and contrast, as has already been presented by \citet{2018A&A...616A..11G}.  We note the well-known trend that stars having high $V$ also have higher $U$.

\subsection{Actions} \label{sec:actions}

\begin{figure}
    \centering
    \includegraphics[width=\columnwidth]{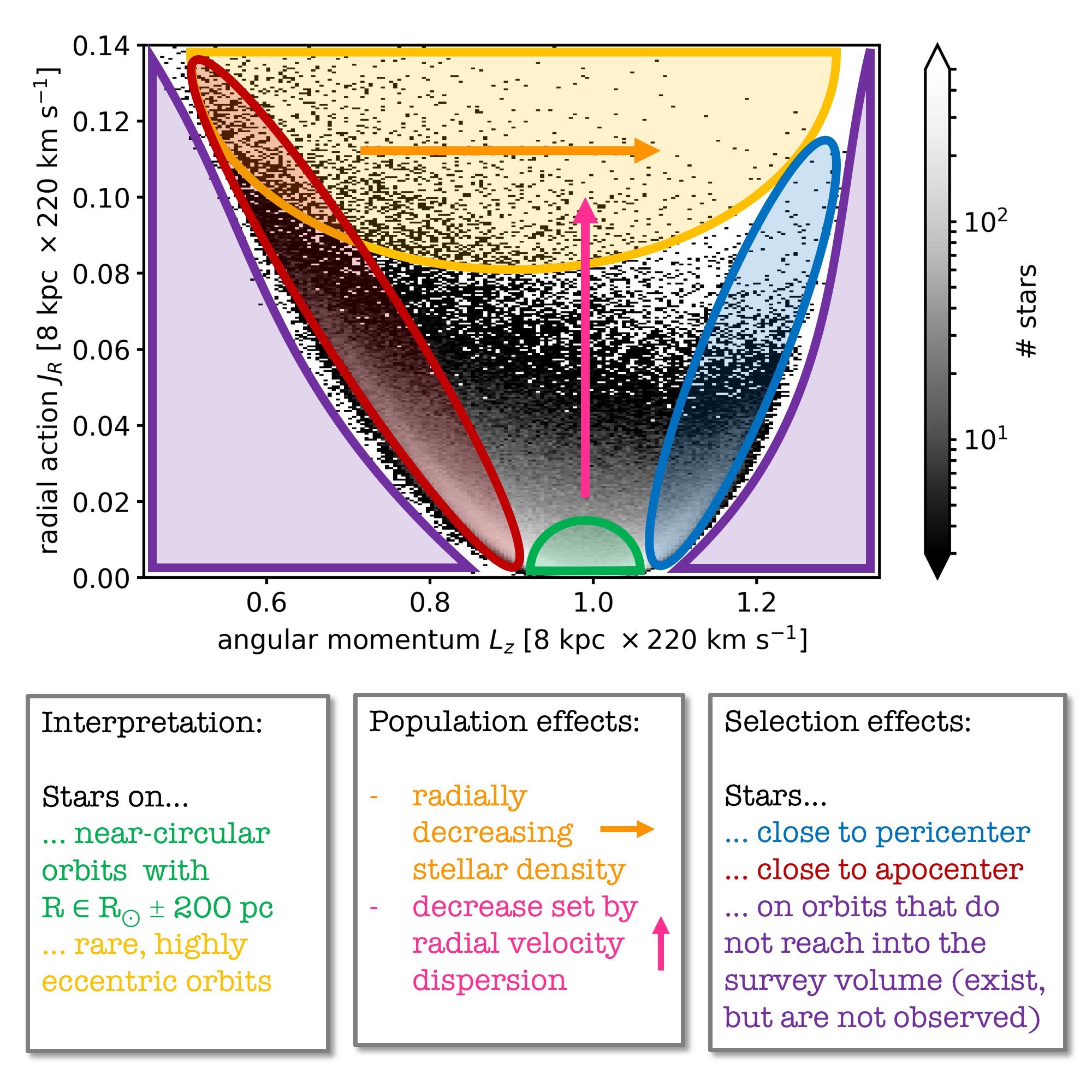}
    \caption{Distribution $n(J_R,L_z)$ of stellar mock data created within a survey volume of $1/\varpi < 200~\text{pc}$ around the Solar position in an axisymmetric, phase-mixed Milky Way disc model (see the text for details). This figure contains $\sim 310,000$ randomly drawn mock stars shown in greyscale in the background of the figure and should be compared to the right-hand panels in Figure \ref{fig:sol_neighbourhood_200pc}. Note the varying, but overall smooth distribution of stars. We have marked different regions in action space $(L_z,J_R)$ that are affected by selection effects due to the restricted survey volume and/or effects of the stellar population in question. This should help the reader to interpret the $(L_z,J_R)$ figures in this work.}
    \label{fig:mock_data}
\end{figure}

\subsubsection{Interpreting actions and angles}

Actions in axisymmetric systems are excellent orbit labels and have an intuitive physical interpretation: for short-axis tube orbits in the Galactic disc, the \emph{radial action} $J_R\in[0,\infty]$ can be considered as a measure of the orbit eccentricity or the radial extent of a disc orbit's in-plane epicyclic rosette. The azimuthal action $J_\phi$ is identical to the \emph{angular momentum} in $z$-direction, $L_z\in[-\infty,\infty]$, and describes the amount of rotation around the Galactic centre. The vertical action $J_z\in[0,\infty]$ quantifies the extent of the vertical excursion of the star around the mid-plane.

The actions can be complemented by a set of angle coordinates $\theta_i$ with $i \in [R,\phi,z]$ to form a set of 6D canonical conjugate coordinates, with $\theta_R$ being related to the location of the star on its epicycle, $\theta_\phi$ the azimuth of the guiding centre with respect to the Sun, and $\theta_z$ the vertical phase. In a fully phase-mixed system, the stars are uniformly distributed in $\theta_i \in [0,2\pi]$.

In a galaxy that exhibits non-axisymmetric structures like bars and spiral arms, $(J_R,L_z,J_z)$ are not fully conserved anymore. $L_z$, for example, will be subject to radial migration \citep{2002MNRAS.336..785S}. As shown by \citet{2018MNRAS.474.2706B}, perturbation theory would need to be invoked to recover the orbit's time evolution in the action-angle framework. At a given point in time, approximations for the current axisymmetric actions can, however, still be calculated.

For an in-depth introduction to action-angle coordinates, we refer the reader to \S3.5 in \citet{2008gady.book.....B}. A heuristic introduction to actions can also be found in \S1.4 of \citet{phdthesis}.

\subsubsection{Action estimation and Milky Way potential} \label{sec:action_estimation_potential}

The exact calculation of actions from the observed $(R,z,v_{R},v_{T},v_{z})$ coordinates of a star is computationally expensive and requires knowledge of the gravitational potential. \citet{2016MNRAS.457.2107S} review different algorithms to efficiently estimate the actions. In this work, we use the \emph{St\"{a}ckel Fudge} estimation algorithm by \citet{2012MNRAS.426.1324B}, of which an implementation is provided in the \texttt{galpy} python package for Galactic Dynamics by \citet{2015ApJS..216...29B}.\footnote{As focal length of the underlying St\"{a}ckel potential's $(u,\nu)$ coordinate system, we use $\Delta = 0.45 \times R_\odot$ \citep{2012MNRAS.426.1324B,2013ApJ...779..115B} for speed.}

\texttt{galpy} also provides an axisymmetric gravitational potential model called \texttt{MWPotential2014} with a power-law bulge, a Miyamoto-Nagai disc, and NFW halo, whose parameters were found from a fit to dynamical data of the Milky Way (for details see \citealt{2015ApJS..216...29B}). At $R_\odot=8~\text{kpc}$ this potential model has a circular velocity of $v_\text{circ}(R_\odot) = 220~\text{km s}^{-1}$. For the 3.6 Million stars in our sample, the action estimation in this potential takes just $\sim8$ minutes on one core.

\subsection{Axisymmetric mock data and selection effects} \label{sec:mock_data}

To illustrate how the distribution of stars in action space, $n(L_z,J_R)$, would look like in the idealized case of a perfectly axisymmetric and phase-mixed Milky Way disc, we create a mock stellar distribution, which we show and interpret in Figure \ref{fig:mock_data}. Here, we also help the reader to understand the selection effects of restricting the observations to a sphere around the Sun. Similar phase-mixed data are also presented in fig. 10 by \citet{2011MNRAS.418.1565M}.

\begin{table*}
    \centering
    \begin{tabular}{ccclll}
        Estimated slope & \multicolumn{2}{c}{Location} & Related to & \multicolumn{2}{c}{Marked in plots with}\\
         $\Delta J_R / \Delta L_z$ & $L_{z,\text{ref}}$ [$L_{z,0}$] & $J_{R,\text{ref}}$ [$L_{z,0}$] & moving group & letter & colour \\\hline\hline
         -0.7 & 0.675 & 0.05 & & \texttt{A} & \texttt{grey}\\
         -0.7 & 0.747 & 0.05 & & \texttt{B} & \texttt{petrol}\\
         -0.8 & 0.835 & 0.05 & & \texttt{C1} & \texttt{green}\\
         -0.8 & 0.884 & 0.05 & & \texttt{D1} & \texttt{blue}\\
         -0.9 & 1.013 & 0.05 & & \texttt{F1} & \texttt{red}\\
         -0.9 & 1.075 & 0.05 & & \texttt{G1} & \texttt{orange}\\
         -0.9 & 1.26 & 0.05 & & \texttt{I} & \texttt{yellow}\\\hline
         -1.8 & 0.837 & 0.01 & Hercules & \texttt{C2} & \texttt{dark green}\\
         0.6 & 0.87 & 0.01 & Hercules & \texttt{D2} & \texttt{dark blue}\\
         -0.5 & 0.978 & 0.01 & Hyades & \texttt{E1} & \texttt{pink}\\
         -0.15 & 0.953 & 0.01 & Hyades & \texttt{E2} & \texttt{purple}\\
         -0.5 & 1.053 & 0.01 & Sirius & \texttt{F2} & \texttt{dark red}\\
         -0.5 & 1.123 & 0.01 & Sirius & \texttt{G2} & \texttt{dark orange}\\
         -0.9 & 1.209 & 0.01 & & \texttt{H} & \texttt{gold}\\\hline
    \end{tabular}
    \caption{Estimated slopes and locations of the lines in the action distribution $n(L_z,J_R)$ at which the \emph{Gaia} DR2/RVS stars exhibit overdensities. The actions were calculated assuming the \texttt{MWPotential2014} \citep{2015ApJS..216...29B} model for the Galaxy's gravitational potential. The lines follow $J_R(L_z) = \left(\Delta J_R / \Delta L_z \right)\times \left( L_z - L_{z,\text{ref}}\right)+ J_{R,\text{ref}}$ and the actions are given in units of $L_{z,0}=8~\text{kpc} \times 220~\text{km s}^{-1}$. We estimate that the uncertainty in the slopes determined by eye can be up to 20\%; a conservative guess on the uncertainty on the location $L_z$ is $0.01L_{z,0}$. The upper part of the tables contains the ridges at high $J_R$. The lines in the lower part that show up only at low $J_R$ are related to the known moving groups.}
    \label{tab:extended_orbit_substructure}
\end{table*}

\subsubsection{Axisymmetric mock data generated from a smooth DF}

The mock data are created via Monte Carlo sampling of a stellar distribution function (DF), following the procedure described in \citet{2016ApJ...830...97T}, Appendix A. The Milky Way model consists of (a) the \texttt{MWPotential2014} potential by \citet{2015ApJS..216...29B} (see Section \ref{sec:action_estimation_potential}), (b) an axisymmetric action-based stellar disc DF, the quasi-isothermal DF\footnote{The scale parameters of the quasi-isothermal DF were chosen to be: disc scale length $h_R^\text{qdf} = 2.5~\text{kpc}$, velocity dispersion $\sigma_{R,0}^\text{qdf}=33~\text{km s}^{-1}$ and $\sigma_{z,0}^\text{qdf}=25~\text{km s}^{-1}$, and exponential velocity scale length, $h_{\sigma,R}^\text{qdf} = 8~\text{kpc}$ and $h_{\sigma,z}^\text{qdf} = 7~\text{kpc}$ \citep{2013ApJ...779..115B}.} by \citet{2011MNRAS.413.1889B}, and (c) a spherical selection function with sharp cut-off at $d=200~\text{pc}$ around the Sun, as described and used in \citet{2016ApJ...830...97T,2017ApJ...839...61T}. We drew $\sim 310,000$ star particles from this DF and selection function. The Galactocentric radial and vertical velocity dispersions of all star particles in this mock sample are $\sigma_{R} = 34~\text{km s}^{-1}$ and $\sigma_{z} = 22~\text{km s}^{-1}$, and are therefore close to the velocity dispersion of the \emph{Gaia} DR2/RVS data within $200~\text{pc}$, $\sigma_{R} = 37~\text{km s}^{-1}$ and $\sigma_{z} = 20~\text{km s}^{-1}$.

The distribution of mock stars in Figure \ref{fig:mock_data} is---apart from number gradients and selection effects as described below---smooth in the $(L_z,J_R)$ plane.

\begin{figure*}
\centering
{\includegraphics[width=\textwidth]{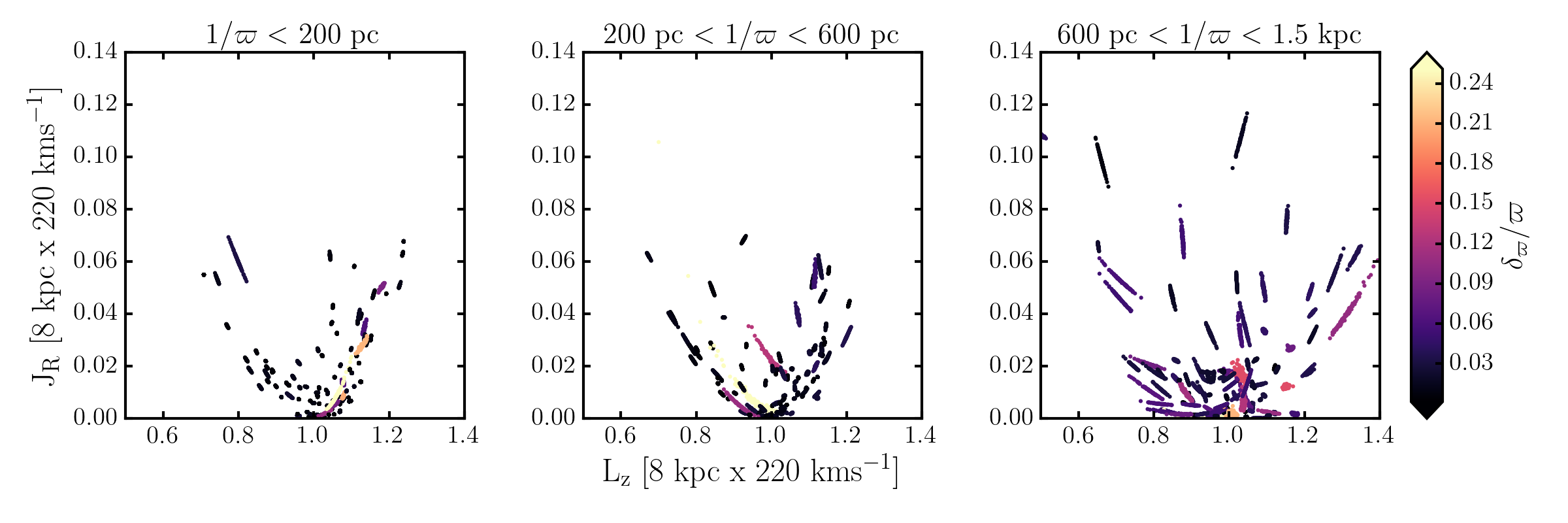}}
\caption{The effect of measurement uncertainties on the estimates of orbital actions. 100 Monte Carlo samples of the \emph{Gaia} DR2/RVS measurement uncertainty ellipses of 100 random stars transformed to action space for the same distance regimes as the ones shown in Figure \ref{fig:sol_neighbourhood}, colour coded by the fractional error in parallax ($\delta_{\varpi}/\varpi$). This figure illustrates that in the era of \emph{Gaia} the measurement uncertainties generally result in small action uncertainties (cf. \citealt{2018MNRAS.481.2970C}). But the figure also illustrates that for stars further away the \emph{Gaia} uncertainties are not negligible, and presumably responsible for the smearing out of some of the structure in Figures \ref{fig:sol_neighbourhood_600pc}-\ref{fig:sol_neighbourhood_1500pc}.}
\label{fig:error_ellipses}
\end{figure*}

\subsubsection{Interpreting the distribution of disc stars in action space} \label{sec:survey_volume}

For the mock action distribution $n(L_z,J_R)$ in Figure \ref{fig:mock_data}, we point out (a) interpretation guidelines, (b) effects of the stellar population, i.e., of the choice of disc DF and its parameters, and (c) selection effects of the limited survey volume.

(a) Stars with $J_R\sim0$ move on near-circular orbits (green region); stars with large $J_R$ on highly eccentric orbits (yellow region).

(b) The decrease in the number of stars towards high $J_R$ (pink) and high $L_z$ (orange) is set by the stellar disc DF we are using. See \citet{2011MNRAS.413.1889B} (and also \citealt{2013ApJ...779..115B}) for details and the exact functional form of the quasi-isothermal DF. As a very rough guideline, 
\begin{equation}
\text{DF}(L_z,J_R) \appropto \exp\left(-\frac{R_g(L_z)}{h_R}\right) \times \exp\left( - \frac{ \kappa J_R}{\sigma_R} \right) \times \exp\left( - \frac{ \nu J_z}{\sigma_z} \right),\label{eq:qdf_guide}
\end{equation}
where $\kappa$ and $\nu$ are the epicycle and vertical frequency, respectively, $h_R$ approximately the disc scale length, and $R_g$ the orbit's guiding-centre radius. For a flat rotation curve, $R_g = L_z/v_\text{circ}=R\times v_T/v_\text{circ}$. The larger the radial velocity dispersion $\sigma_R$ of the population, the more stars are on high-eccentricity orbits at large $J_R$. The $R_g(L_z)$-term reflects the radial exponential decrease in stellar density of the disc model.

(c) The parabolic envelope at the lower edge of $n(L_z,J_R)$ reflects the radial extent of the sample's spherical selection function. Stars within $\Delta R$ of the Sun on perfectly circular orbits, $J_R=0$, can only be observed if their $L_z$ is in the range
\begin{equation}
L_{z,0} \pm \Delta L_z \simeq \left(R_\odot\pm \Delta R\right) \times v_\text{circ}
\end{equation}
with $L_{z,0} \equiv 8\text{kpc}\times 220\text{km s}^{-1}$. This is the extent in $L_z$ at $J_R=0$ covered by the green region in Figure \ref{fig:mock_data}, where $\Delta R = 200~\text{pc}$ (and the green horizontal bars in Figures \ref{fig:sol_neighbourhood} and \ref{fig:action_distribution_inside_outside} in the subsequent sections for their respective $\Delta R$). Stars with $L_z$ outside of this range, i.e., stars that live on average further away from the Sun, cannot enter the survey volume if they have low $J_R$ (purple region). Stars located near the low-$J_R$ edge of the distribution are on orbits that are just eccentric enough given their $L_z$ to be able to reach into the survey volume. The larger $|L_z - L_{z,0}|$, the larger $J_R$ has to be. Stars observed at $L_z/L_{z,0} < 1$ near the low-$J_R$ edge are therefore currently close to apo-centre (red region); stars at the $L_z/L_{z,0} > 1$, low-$J_R$ edge are close to peri-centre (blue region). 

The parabolic envelope can be explained with Equation (3.261) in \citet{2008gady.book.....B} which describes a star's radial oscillation around its $R_g$ (in the epicyclic approximation),
\begin{equation}
R(t) = R_g - \sqrt{\frac{2 J_R}{\kappa}} \cos \theta_R(t),
\end{equation}
where $\kappa$ is the epicycle frequency. For a flat rotation curve, the action distribution in the plane falls therefore into the parabolic envelope
\begin{equation}
J_R(L_z) > \min\left\{\frac{\kappa}{2} \left(L_z/v_\text{circ} - R\right)^2 \mid R \in \text{survey volume}\right\}.
\end{equation}

\section{Results}

We are now in a position to interpret the actual distribution of $n(L_z,J_R)$ for subsamples of the \emph{Gaia} DR2/RVS stars and compare it to our simplistic expectations. We do this by both considering different subsamples in Section \ref{sec:extended_orbit_structure}, and by comparison to the substructure in velocity space found by \citet{2018A&A...616A..11G} in Section \ref{sec:moving_groups}. We  will conclude with some investigations on the asymmetric distribution of radial velocities (Section \ref{sec:asymmetric_vR}) and the vertical action $J_z$ (Section \ref{sec:mean_Jz}).

\subsection{The extended orbit structure of the Galactic disc within 1.5 kpc} \label{sec:extended_orbit_structure}

\subsubsection{Substructure along linear lines in action space}

In the middle and right-hand panels of Figure \ref{fig:sol_neighbourhood}, we plot for the distance bins motivated in Appendix \ref{sec:data_quality} 2D histograms of the number of stars in action space, in particular, in the actions that quantify the oscillations in the plane of the disc, $L_z$ and $J_R$. This shows for the first time the rich substructure in the space of orbital actions out to $\sim1.5~\text{kpc}$. Comparison to the axisymmetric mock data in Figure \ref{fig:mock_data} shows a highly structured distribution with several ridge-like overdensities in $n(L_z,J_R)$ extending towards high $J_R$. Many of these ridges appear to lie along lines with constant $\Delta J_R /\Delta L_z$. They are present over two orders of magnitudes in number density, which makes an assumption-free, algorithmic selection, with e.g. a $k$-mean-lines algorithm, difficult. We have therefore located stellar overdensities by eye that show up in all distance bins. In Panel \ref{fig:sol_neighbourhood_600pc} for $200~\text{pc} < 1/\varpi < 600~\text{pc}$, they are best visible: we have counted seven ridges at high $J_R$ and substructure in seven clumps at low $J_R$. We have estimated the slopes of the 14 overdensities by eye and summarized their values in Table \ref{tab:extended_orbit_substructure}. In the right-hand panels of Figure \ref{fig:sol_neighbourhood}, we overplot these lines on top of the action distribution. Even though the substructure seems to blur out (see Section \ref{sec:measurement_uncertainties}),  we  find  that  several  of  the  overdensities  are located at the same positions in $(L_z,J_R)$ as in the local sample $1/\varpi < 200~\text{pc}$. This demonstrates that it is truly the same extended orbital substructure of the disc that we see everywhere out to $~1.5~\text{kpc}$.

We note the different behaviour at $J_R$ smaller and larger than $\sim 0.05 L_{z,0}$. At high $J_R$, the seven identified features all exhibit negative slopes of $\Delta J_R / \Delta L_z \sim 0.7-0.9$ over a large $\Delta J_R$, and most of the ridges appear to occur in close, parallel pairs. At $J_R \lesssim 0.05L_{z,0}$, the apparent continuations of the high-$J_R$ features have different slopes; either because the features might be weakly curved rather than perfectly linear over the whole $J_R$-range, or because the features fan out towards higher $J_R$ (which can be best seen at $1/\varpi < 600~\text{pc}$ for the feature marked in \texttt{green (C)} in Table \ref{tab:extended_orbit_substructure} and Figure \ref{fig:sol_neighbourhood_1500pc}), or they are truly independent features. We found one feature with a positive slope (\texttt{dark blue, D2}), and one feature that extends to a particular prominent low-density region at high $J_R$ (\texttt{gold, H}). This gap, as well as the gap between the \texttt{petrol (B)} and \texttt{green (C1)} features, was visible even in the blurred-out action distribution of \emph{Gaia} DR2/RVS stars beyond $1.5~\text{kpc}$ out to $3~\text{kpc}$.

\begin{figure*}
\centering
\includegraphics[width=\textwidth]{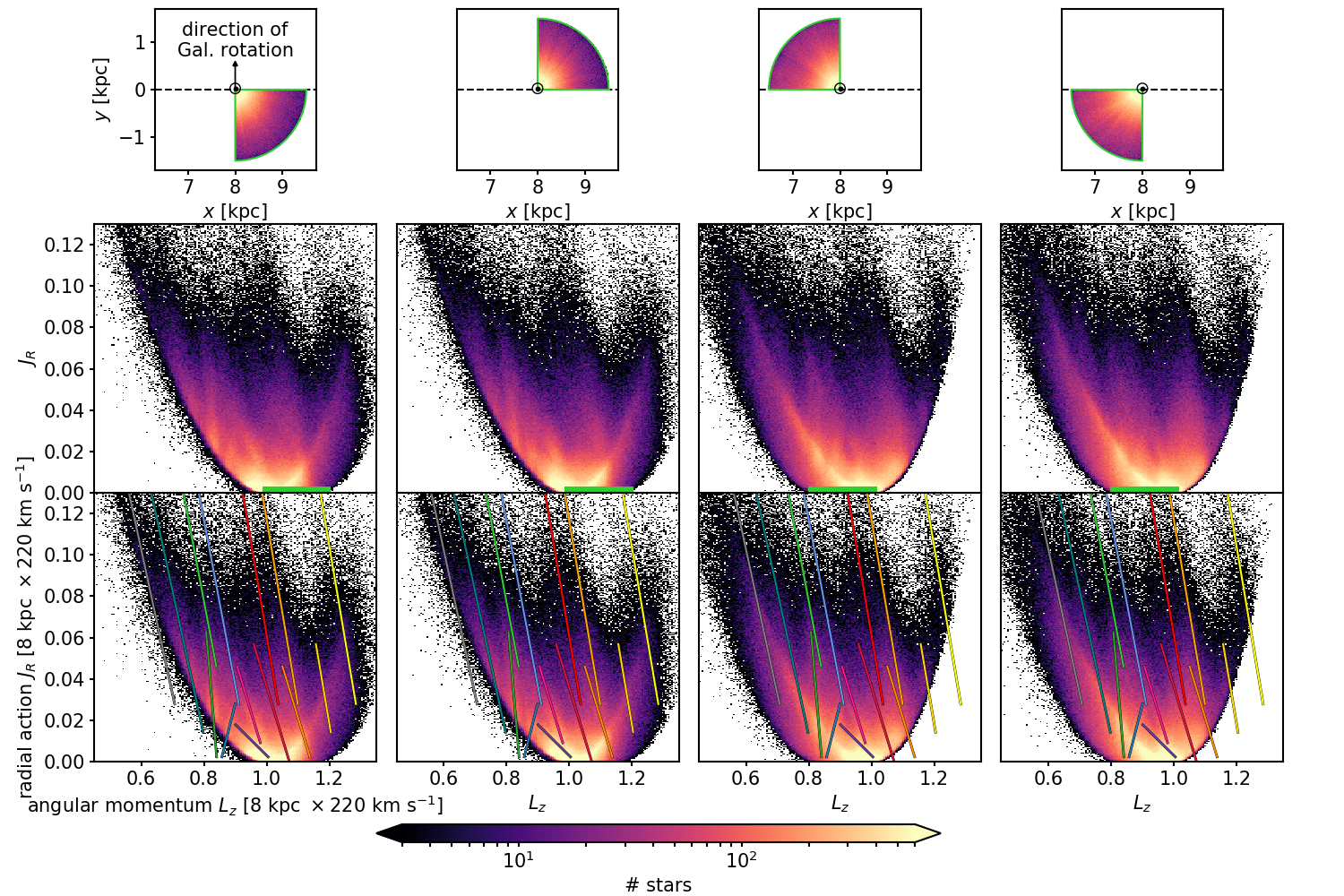}
\caption{Distribution $n(L_z,J_R)$ of the high-quality \emph{Gaia} DR2/RVS sample within $1/\varpi<1.5~\text{kpc}$ from the Sun, for different sectors in the Galactocentric $(x,y)$ plane centred on the Solar position ($\odot$ at $(x,y) = (8,0)~\text{kpc}$), as indicated in the upper row. We overplot the overdensity lines summarized in Table \ref{tab:extended_orbit_substructure} in the lower row; they are identical for all four sectors. This figure illustrates that the extended orbit structure shows up at the same locations in $(L_z,J_R)$ space and with the same slopes both inside and outside of the Sun's position, and in front or behind the Sun in the direction of Galactic rotation. The relative strength of the features, however, might vary from location to location. The green horizontal bar in the middle panels denotes the $L_z$-range corresponding to the radial coverage of each subsample, and the parabolic lower edges as well as the parabolic stellar overdensities emanating from $L_z/L_{z,0}=1$ are selection effects; the latter in particular is caused by the sharp vertical edge and the large number of stars in \emph{Gaia} DR2 at $R=8~\text{kpc}$.}
\label{fig:action_distribution_inside_outside}
\end{figure*}

\subsubsection{The influence of measurement uncertainties} \label{sec:measurement_uncertainties}

It is necessary to explore to which extent the measurements in the  observables ($\varpi , \vec{\mu}, v_\text{los}$) propagate into the space of $(L_z,J_R)$. In Figure \ref{fig:error_ellipses}, we illustrate the extent of the morphology in $n(L_z,J_R)$ that could be affected by measurement uncertainties. For 100 random stars in each distance bin, we converted 100 Monte Carlo samples each drawn from the \emph{Gaia} uncertainty ellipse (including the covariances) into action space. The majority of stars at close distances ($1/\varpi <200~\text{pc}$) have a small error in parallax. We also note that the uncertainty distributions are much narrower with respect to stars at larger distances where the uncertainties tend to spread more in action space. This is expected to cause some blurriness in action space that could hide possible structures, in particular in Figure \ref{fig:sol_neighbourhood_1500pc}.

Overall, the uncertainties are much smaller than the observed substructures, especially in the regime $1/\varpi <200~\text{pc}$. So we expect that these action features indeed exist in the Milky Way and are not just observational relics.

\begin{figure*}
    \centering
    \includegraphics[width=\textwidth]{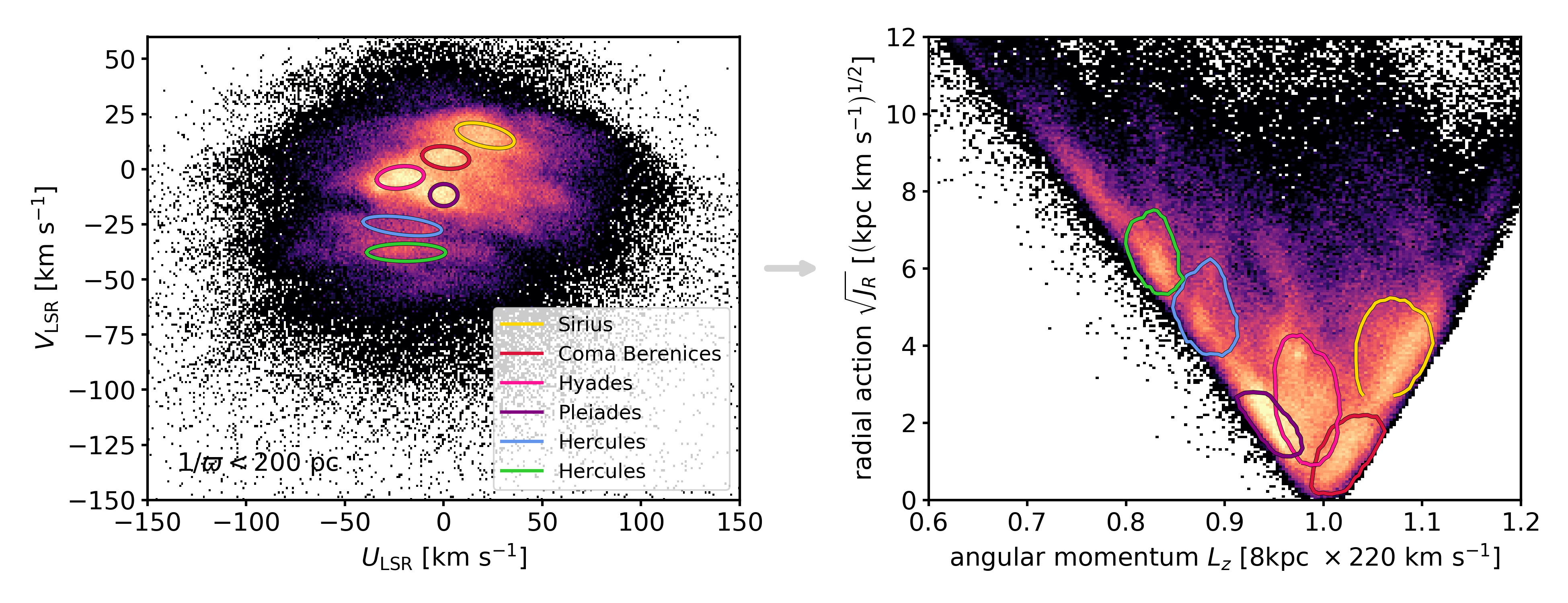}
\caption{Known moving groups in the Solar neighbourhood ($1/\varpi < 200~\text{pc}$) identified in the $(U,V)$ velocity plane of \emph{Gaia} DR2/RVS, and the location of the corresponding stars in the orbital action plane $(L_z,J_R)$. In the left-hand panel---the $(U_\text{LSR},V_\text{LSR})$ plane---overdensities in $n(U,V)$ were picked based on the locations noted in \citet{2018A&A...616A..11G} and marked by approximate ellipses. In the region of the Hercules stream, $(U,V)\sim(-30,-50)~\text{km s}^{-1}$ (\citealt{2008A&A...490..135A}, and references therein), two slightly separate density arches were found by \citet{2018A&A...616A..11G} (\texttt{green} and \texttt{blue}). Stars that fell into each of the ellipses were then located in the action plane in the right-hand panel. The contours contain 95\% of the stars. The known moving groups correspond clearly to the most prominent overdensities in $n(L_z,J_R)$. (Here, we have plotted $\sqrt{J_R}$ rather than $J_R$ to put more visual emphasis on the low-$J_R$ moving groups.) The stars that make up the bulk of the moving groups are (with the exception of the Hyades) confined to the edges of the distribution, which are strongly dominated by selection effects (see Figure \ref{fig:mock_data}). The small bright clump at $\sqrt{J_R}\sim 4~(\text{kpc km/s})^{1/2}$ within the \texttt{pink} contour of the Hyades stream contains the stars of both the Hyades and Praesepe open clusters.}
\label{fig:UV_arrow_action}
\end{figure*}

\subsubsection{The orbit structure at different locations in the Galactic plane}

A second test, to demonstrate that we see here not just local clumps, but indeed a system of orbit features consistently extending over almost $\Delta R=3~\text{kpc}$ in the Galactic disc, is shown in Figure \ref{fig:action_distribution_inside_outside}. Here, we split our total stellar sample into angular wedges around the Sun, inside and outside of the Solar circle at $R=8~\text{kpc}$, and in front and behind the Sun in the direction of Galactic rotation. We overplot the lines from Table \ref{tab:extended_orbit_substructure} onto action space analogously to Figure \ref{fig:sol_neighbourhood}.

The selection effects are here slightly different: Note the location of the green horizontal bar in the central panels marking the radial range covered, and the parabolic overdensities extending from its end point at $L_z/L_{z,0} = 1$, which corresponds to the sharp edge at $R=8~\text{kpc}$ where \emph{Gaia}'s completeness is highest.

The overdensities visible still follow our overplotted lines that all have the same location and slopes in action space for all four different regions in the Galactic plane with different $(R,\phi)$ or $(x,y)$. There might be, however, slight differences in the relative strength of each feature.

A similar investigation for stars at different heights above and below the Galactic plane also resulted in a distribution very similar to Figure \ref{fig:sol_neighbourhood_600pc}, with the overdensities at the same locations (see also Figure \ref{fig:mean_vertical_action}).

\subsection{Comparison with known moving groups in the Solar neighbourhood} \label{sec:moving_groups}

It has long been known that the Solar neighbourhood contains overdensities of stars that move together on the same orbits, originally identified in the $(U,V)$ plane of stars within $200~\text{pc}$ of the Sun \citep{1998AJ....115.2384D}. In Appendix \ref{sec:moving_groups}, we include a small review of the current state of research about these \emph{moving groups} for the interested reader. The commonly accepted picture is that these moving groups have a dynamic origin and are caused by spiral arms and/or bar resonances (see e.g., \citet{2010LNEA....4...13B}, and references therein). In the following, we investigate how the moving groups are related to the substructure in action space.

\subsubsection{The moving groups in action space}

In Figure \ref{fig:UV_arrow_action}, we have marked and labelled the known moving groups in the left-hand panel showing $(U,V)$ space of the local $1/\varpi < 200~\text{pc}$ subsample. We used the peak $(U,V)$ positions recorded by \citet{2018A&A...616A..11G} for the moving groups Sirius, the Hyades and Pleiades, and the two Hercules substreams, and drawn ellipses by eye around them to encompass the most prominent regions of the overdensities. The stars within these ellipses were then mapped into action space. Contours encompassing 95\% of the stars of each group are shown in the same colours in the right-hand panel of Figure \ref{fig:UV_arrow_action} in the $(L_z,\sqrt{J_R})$ plane. (We plot here the distribution in $\sqrt{J_R}$ rather than $J_R$ to visually emphasize the substructure at low $J_R$. The parabolic lower edge becomes a straight line.)

We see now that---with the exception of the Hyades stream---the clumps at the low-$J_R$ edge of the action distribution correspond to the moving groups. This is not surprising, as stars in the local Solar neighbourhood share almost the same $(R,\phi,z)$ position. If they also have similar $(U,V)$ velocities in the disc plane, it is very likely that they are overall on very similar orbits and therefore also have similar actions describing the horizontal motions, i.e. $(L_z,J_R)$.

There appears to be a slight tendency of the groups at low $J_R$ to extend into some of the observed high-$J_R$ ridges. As we explained in Section \ref{fig:mock_data} and Figure \ref{fig:mock_data}, the low-$J_R$ edge of the action distribution is strongly shaped by selection effects. We will look into this further in the next section.

Our Figure \ref{fig:UV_arrow_action} should be compared to fig. 10 in \citet{2010MNRAS.409..145S}, who did the same exercise of mapping moving groups to action space with Hipparcos/GCS data. The most prominent feature in his study was the Hyades stream; other structures were barely visible. This illustrates once more how data from \emph{Gaia} DR2 can and will revolutionize our knowledge about the Galaxy.

\begin{figure*}
    \centering
    \includegraphics[width=\textwidth]{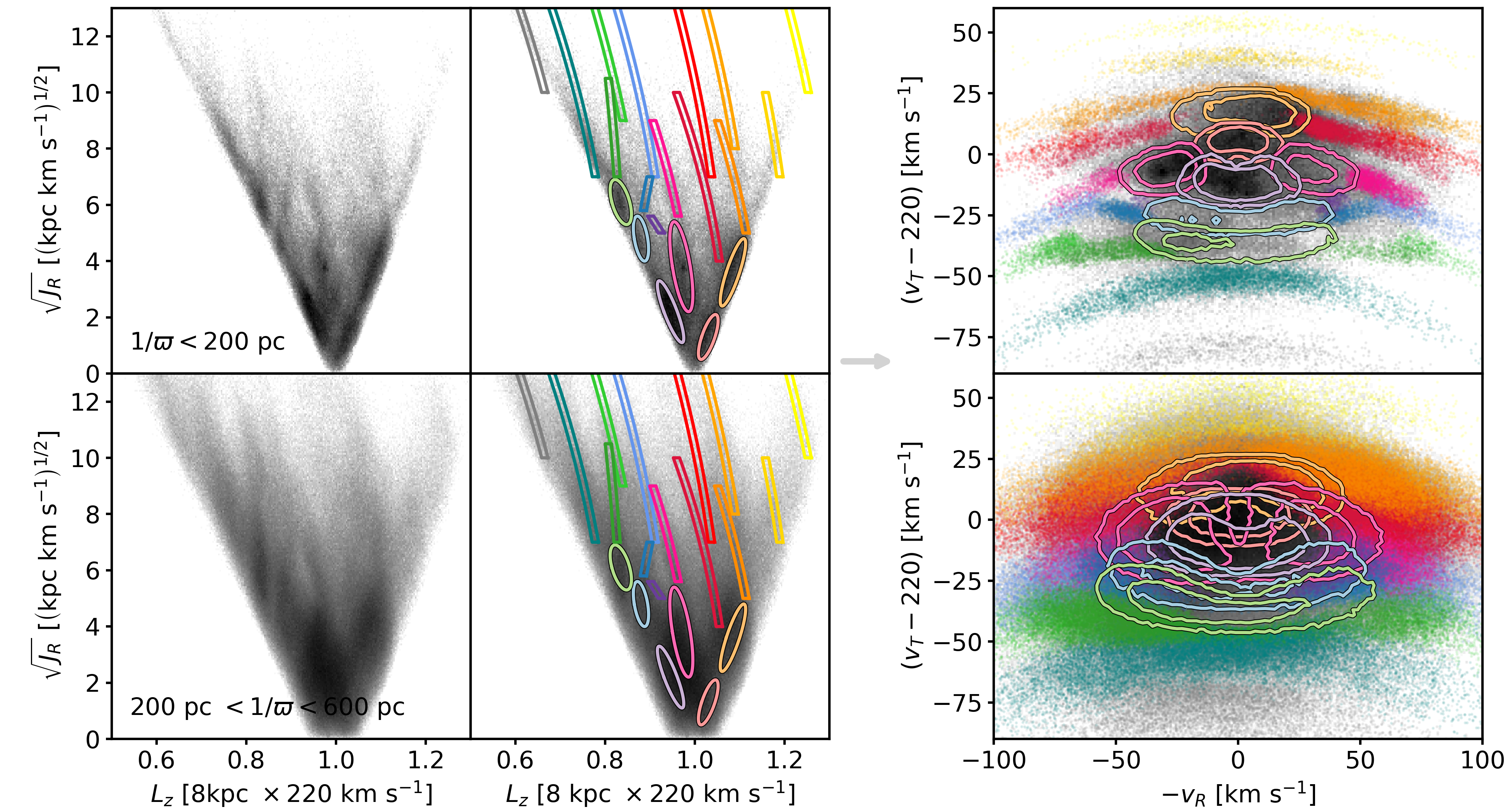}
    \caption{Mapping of regions in action space $(L_z,\sqrt{J_R}$) into the $(-v_R,vT)$ velocity plane, for stars inside of $1/\varpi < 200~\text{pc}$ (upper panels) and $200~\text{pc} < 1/\varpi < 600~\text{pc}$ (lower panels). The \emph{Gaia} DR2/RVS stars are shown in the background with a logarithmic greyscale colour map. In the middle panels, we have overplotted: (a) Boxes of width $\Delta L_z = 0.014L_{z,0}$ around the lines marking the extended orbit structure given in Table \ref{tab:extended_orbit_substructure}, truncated at lower $J_R$. (The lines show up as curves, as we plot $\sqrt{J_R}$ rather than $J_R$ to focus on the low-$J_R$ region.) (b) Ellipses at the location of the moving groups as identified in action space in Figure \ref{fig:UV_arrow_action}. The stars within these boxes and ellipses are then mapped into the velocity plane: (a) the stars of the extended orbit overdensities as coloured points and (b) the stars in the ellipses as coloured contours at the levels of 5 and 100 stars per $(2~\text{km s}^{-1})^2$ bin. This illustrates that: (i) The action overdensities that correspond to the moving groups in the $200~\text{pc}$ sample (upper panels) do not have the same shape, extent, and orientation in the larger $600~\text{pc}$ sample (lower panels). This supports the picture that considers the moving groups as local, and selection-effect-affected manifestation of the extended orbit structure we found in this work. (ii) In the local Solar neighbourhood (upper panels), the overdensity ridges at high $J_R$ are related to the velocity arches found by \citet{2018A&A...616A..11G}, but appear to show up much clearer in action space. (iii) A star sample at the same $(L_z,J_R)$ contains both inward and outward moving stars (as expected), but to different numbers (see also Section \ref{sec:asymmetric_vR}). (iv) Velocity space beyond $200~\text{pc}$ blurs stars that belong to different orbit substructures (right-hand panels). Beyond the immediate solar neighbourhood, it is therefore more informative to investigate orbits in action rather than velocity space.}
    \label{fig:action_arrow_velocity}
\end{figure*}

\subsubsection{Mapping features in $n(L_z,J_R)$ to $n(-v_R,v_T)$}

In the previous section, we saw that within $200~\text{pc}$ from the Sun action space and velocity space are basically equivalent and both contain the moving groups as overdensities. We will now point out some subtle differences and extend the discussion beyond the Solar neighbourhood.

In the left-hand panels of Figure \ref{fig:action_arrow_velocity}, we use the lines from Table \ref{tab:extended_orbit_substructure} to draw boxes of width $\Delta L_z = 0.014L_{z,0}$ around the overdensities of the extended orbit structure in action space. At low $J_R$, we truncate these boxes and add ellipses that mark the location of the moving groups based on the local $1/\varpi < 200~\text{pc}$ sample in the upper panel (cf. Figure \ref{fig:UV_arrow_action}).

The first thing to notice is that the moving group ellipses (in particular Sirius and Hercules) do not follow the orientation and extent of the corresponding overdensities in the $200~\text{pc} < 1/\varpi < 600~\text{pc}$ action distribution anymore. The overdensities rather lie along the lines identified in Figure \ref{fig:sol_neighbourhood} and Table \ref{tab:extended_orbit_substructure}. As mentioned earlier, the reason is that the moving groups are confined to regions in action space of which we know that they are strongly shaped by selection effects. Consequently, the moving groups in velocity space need to be considered always in the context of a selection function.

In the next step, we map the stars within the ellipses and boxes into the $(-v_R,v_T)$ plane (right-hand panels in Figure \ref{fig:action_arrow_velocity}). Stars within the boxes at high $J_R$ are overplotted as colourful dots in velocity space; stars within the ellipses at low $J_R$ are marked with contours drawn at 5 and 100 stars per $(2~\text{km s}^{-1})^{2}$ bin. Obviously, both stars moving towards the Galactic centre $(-v_R \gg 0)$, as well as away from the Galactic centre $(-v_R \ll 0)$, contribute to the same feature in action space---albeit with different numbers of stars. We will look into this in more detail in Section \ref{sec:asymmetric_vR}.

For the local $1/\varpi < 200~\text{pc}$ sample, the high-$J_R$ ridges in $n(L_z,J_R)$, starting at the location of known moving groups and extending towards high $J_R$, turn out to map to the arch-like extensions at high $|v_R|$ that were observed by \citet{2018A&A...616A..11G}. We claim here, however, that these features and their orientation are best identified in action space, as the ridges in action space are tightly concentrated along linear lines, as opposed to be located within extended arches in velocity space. The gaps in the Hercules stream---in Figure \ref{fig:action_arrow_velocity} the gaps between \texttt{petrol (B)}, \texttt{green (C)}, and \texttt{blue (D)}---were discovered by the \citet{2018A&A...616A..11G}. Action space reveals another weak gap that separates the \texttt{grey (A)} and \texttt{petrol (B)} feature. It appears that the strongest gap is, however, the one between \texttt{orange (G)} and \texttt{yellow (I)}, which in velocity space looks like an arch-like boundary at $V_\text{LSR}\sim25~\text{km s}^{-1}$.

At distances beyond $200~\text{pc}$, there is still a plethora of structure discernible in action space, while the velocity distribution becomes quite rapidly ``blurry'' and indistinct. Even though similar trends as in the local sample in the upper panels can be seen, the different orbit features are strongly overlapping in the velocity plane the further away from the Sun we go (lower panels in Figure \ref{fig:action_arrow_velocity}). The reason is that velocities beyond a tiny volume around the Sun cannot be used as globally valid orbit labels anymore. Action space outside of the Solar neighbourhood might therefore be better for investigating orbit structure than velocity space (see also \citet{2013MNRAS.430.3276M} and the discussion in Section \ref{sec:discussion_orbit_structure}.)

\begin{figure*}
    \centering
    \includegraphics[width=\textwidth]{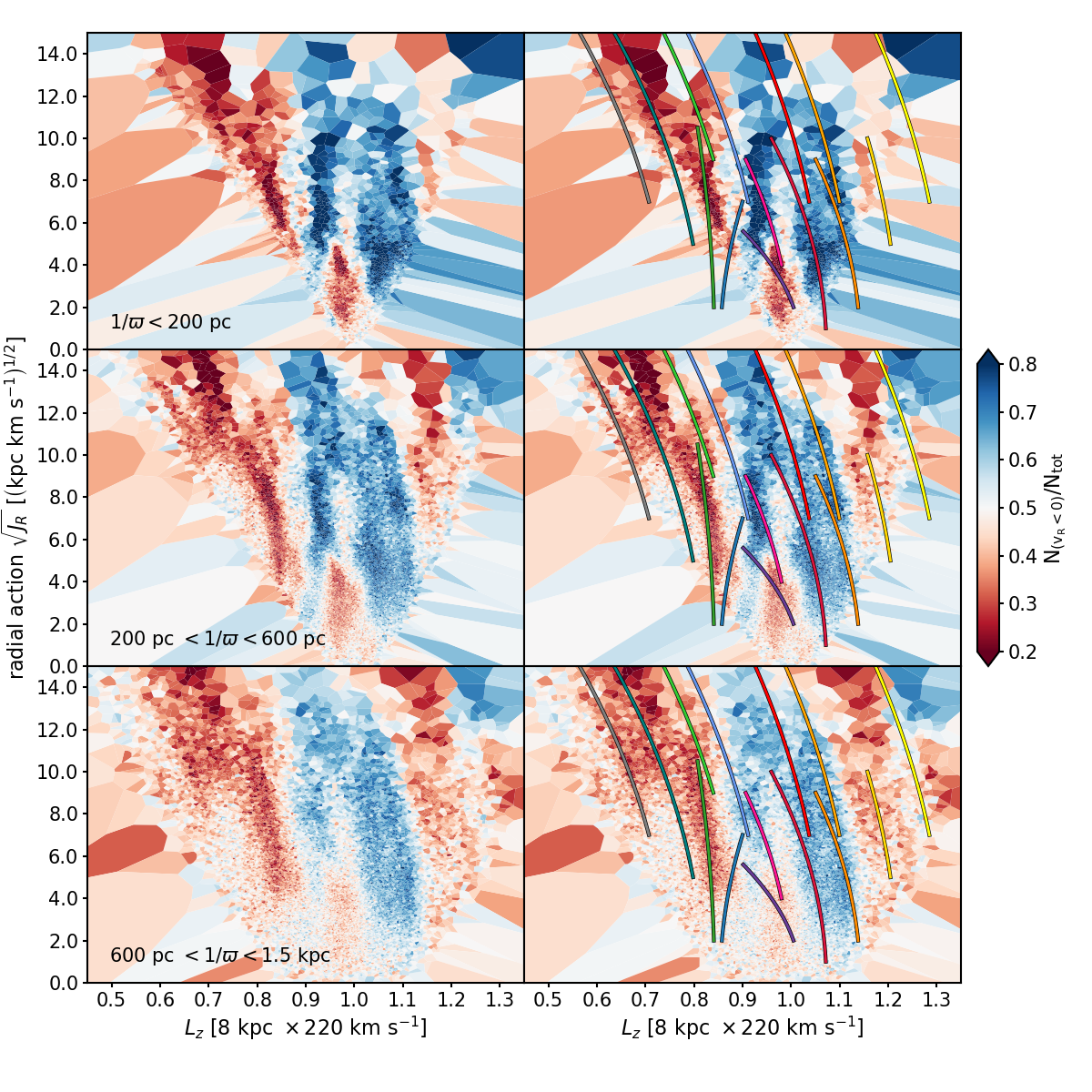}
    \caption{In-out asymmetry in the radial motions as a function of the actions $(L_z,\sqrt{J_R})$, illustrated through the fraction of stars that move radially inwards $v_R < 0$ in a Voronoi tessellation, each containing 100 stars. Blue means that more stars are moving in the direction of the Galactic centre; red means that more stars are moving radially outwards towards the Galactic anti-centre; white means that stars are symmetrically distributed in $\text{sign}(v_R)$, as expected from an axisymmetric Galaxy model. The left- and right-hand panels are identical, except that we have overplotted the loci of the extended orbit structure from Table \ref{tab:extended_orbit_substructure} on the right-hand panels to demonstrate that (i) many, but not all, regions of high density identified by eye are related to regions of highly asymmetric distributions in $v_R$, and (ii) the asymmetry patterns remain overall the same from the Solar neighbourhood out to $\sim 1.5~\text{kpc}$.}
    \label{fig:voronoi}
\end{figure*}

\subsection{Asymmetric population of stars along the orbits} \label{sec:asymmetric_vR}

In a completely phase-mixed, axisymmetric galaxy, one would expect that the stars along non-resonant orbits are evenly distributed in the orbital phases (i.e., the angles $\theta_i$). An eccentric orbit crossing the Solar neighbourhood should then have the same number of stars moving outward and inward.
This need not hold true if the orbital angles are not uniformly populated, or if it is a resonant orbit.

In Figure \ref{fig:action_arrow_velocity}, we saw that the number densities of stars on similar orbits are asymmetrically distributed in $v_R$. This has of course been noted before by several authors (see e.g., \citealt{2015ApJ...800L..32A}, and references therein). Stars at $L_z/L_{z,0} > 1$ and large $J_R$ are more likely to have negative $v_R$ velocities, i.e., they move from the outer Galaxy towards the inner Galaxy (i.e. the Sirius stream and its extensions in \texttt{orange} and \texttt{red}). Vice versa, stars at $L_z/L_{z,0} < 1$, in particular, the Hyades (\texttt{pink}) and Hercules streams with extensions (\texttt{green/blue}), live on average in the inner Galaxy and are currently more likely to move outwards.

To illustrate this further, we have in Figure \ref{fig:voronoi} colour coded a Voronoi tessellation of action space $(L_z,\sqrt{J_R})$ with each 100 stars per bin by the fraction of stars moving towards the Galactic centre, with red for predominant outward motions, and blue for inward motions. We find an asymmetry pattern with sloped, almost vertical stripes, and of the four strongest ones in the Solar neighbourhood one appears red and three appear blue. This asymmetry pattern remains qualitatively the same for all three distance bins out to $\sim1.5~\text{kpc}$ that we consider. In Figure \ref{fig:voronoi}, we have also overplotted the loci of the extended orbit structure from Table \ref{tab:extended_orbit_substructure}. It appears that almost all of the overdense action features are related to regions of high asymmetric motions.

The regions of the moving groups Coma Berenices and the Pleiades (located left and below the \texttt{purple (E2)} feature) appear to be quite phase-mixed in $v_R$. The other moving groups aligned with the low-$J_R$ edge in Figure \ref{fig:UV_arrow_action} (Hercules, Sirius) have no similarly shaped counterparts in the $v_R$-asymmetry pattern in the $1/\varpi < 200~\text{pc}$ sample (upper panels in Figure \ref{fig:voronoi}). In fact, the orientation of the stripy asymmetry patterns is independent of the selection function edge in all volumes.

\begin{figure*}
    \centering
\subfigure[$\sim1.9$ Million stars within $|z| < 150~\text{pc}$ and $1/\varpi < 1.5~\text{kpc}$. \label{fig:mean_vertical_action_S}]{\includegraphics[width=0.9\textwidth]{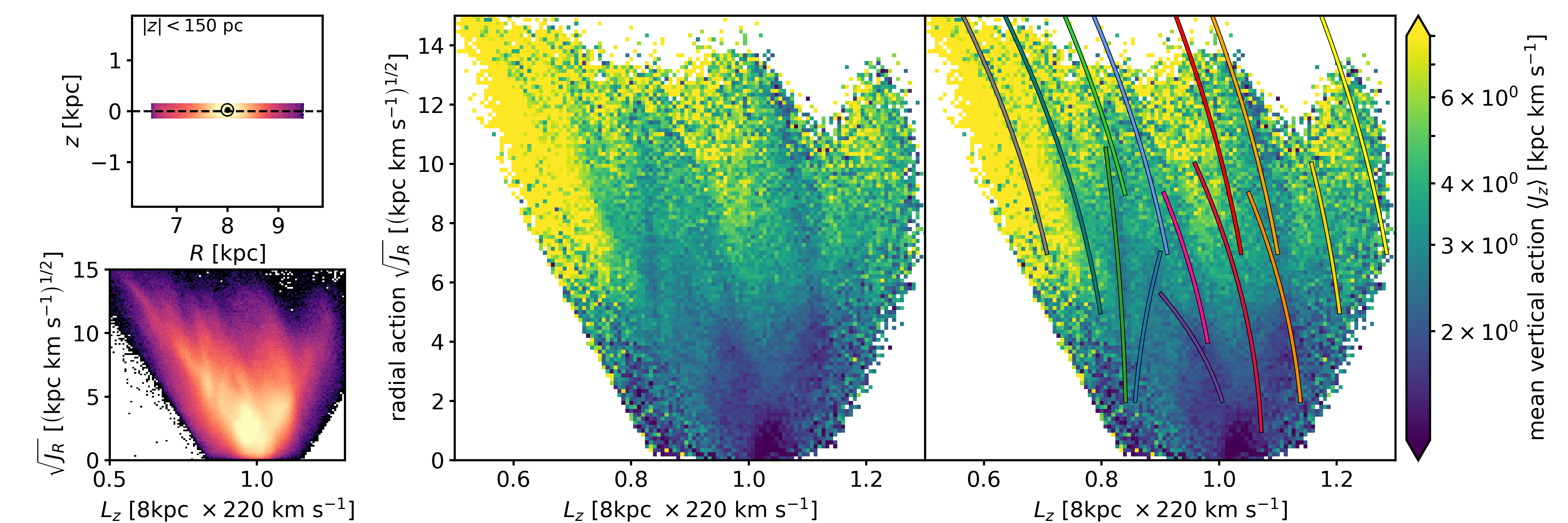}} %
\subfigure[$\sim1.7$ Million stars within $150~\text{pc} < |z| < 500~\text{pc}$ and $1/\varpi < 1.5~\text{kpc}$. \label{fig:mean_vertical_action_M}]{\includegraphics[width=0.9\textwidth]{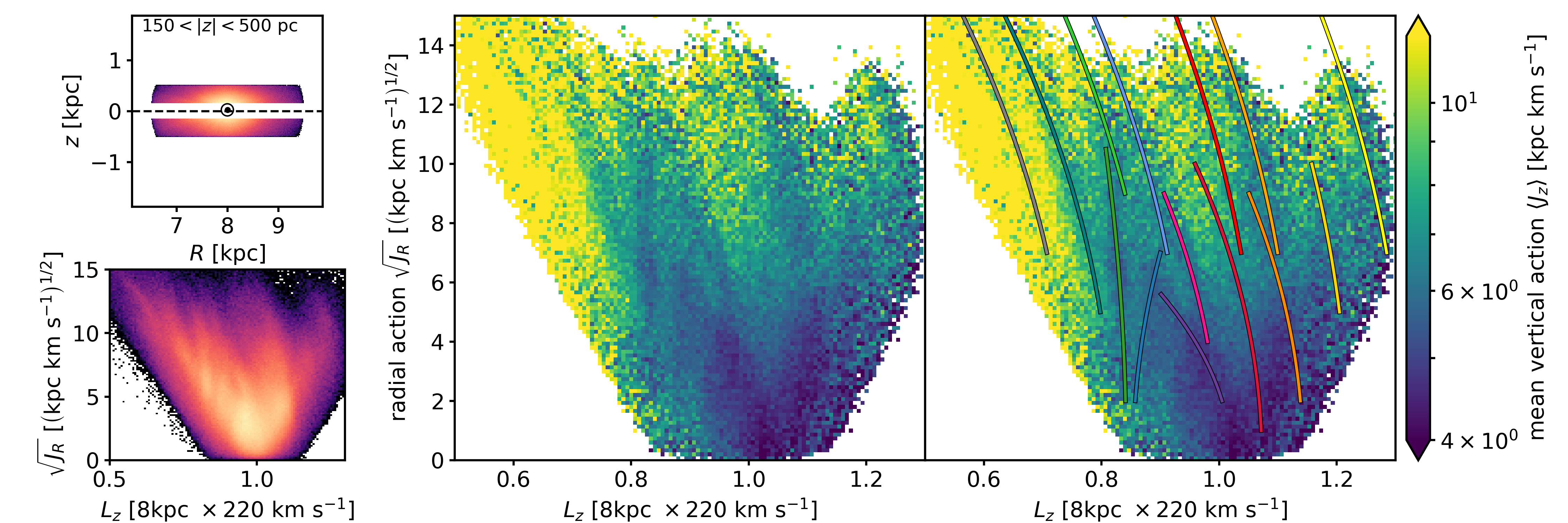}} %
\subfigure[$\sim270,000$ stars within $|z| > 500~\text{pc}$ and $1/\varpi < 1.5~\text{kpc}$. \label{fig:mean_vertical_action_L}]{\includegraphics[width=0.9\textwidth]{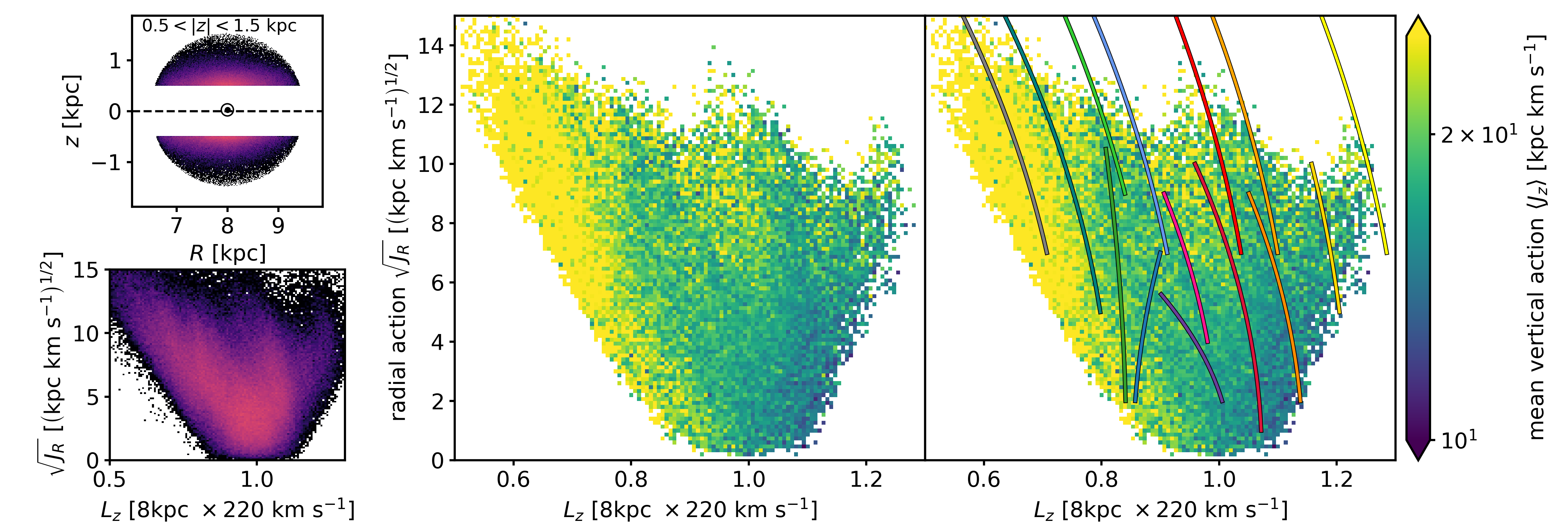}} %
    \caption{The mean vertical action $J_z$ of stars falling into the same bin in $(L_z,\sqrt{J_R})$ space, for three different slices in height away from the Galactic plane, $|z|$, as indicated in the upper left panels. The lower left panels show the corresponding $n(L_z,\sqrt{J_R})$ density distribution for each sample. In the panels on the right-hand side, we overplot the location of the extended orbit structure from Table \ref{tab:extended_orbit_substructure}. Note that the range of the colour bars is different, as stars further away from the disc plane have in general higher $J_z$. Bins have widths $0.13[\text{kpc km s}^{-1} ]^{1/2}$ in $\sqrt{J_R}$ and $0.0067 L_{z,0}$ in $L_z$ and are only plotted if they contain more than 10 stars. This figure illustrates that the substructures in $n(L_z,J_R)$, investigated in the previous figures, are not just stellar overdensities, but also have on average lower vertical actions than the stars in the same $|z|$ range. Only for $|z| > 500~\text{pc}$ the substructure becomes less pronounced both in number density and in the vertical action.}
    \label{fig:mean_vertical_action}
\end{figure*}

\subsection{Features in $n(L_z,J_R)$ and their vertical action} \label{sec:mean_Jz}

So far, we have restricted our investigations of substructure in action space on the actions that govern the motion in the plane of the disc, $L_z$ (circular motion) and $J_R$ (radial motion). In the epicyclic approximation---which is in general valid for most orbits in the thin disc---the radial and vertical motions are decoupled from each other, and structure in the plane of the disc does not necessarily translate into structure in the vertical direction. Radial migration, for example, is expected to change $L_z$ \citep{2002MNRAS.336..785S} and $J_R$, but conserve the vertical action $J_z$ on average \citep{2012MNRAS.422.1363S,2016ApJ...824...39V}.

To investigate substructure in vertical orbit space, we calculate for each bin in the $(L_z,\sqrt{J_R})$ plane ($\Delta \sqrt{J_R} = 0.13 [\text{kpc km s}^{-1} ]^{1/2}$, and $\Delta L_z = 0.0067 L_{z,0}$) the mean vertical action $\langle J_z \rangle$ of the stars and show the result in Figure \ref{fig:mean_vertical_action} for three slices in the $|z|$-coordinate away from the Galactic plane. We overplot the lines of the extended orbit structure from Table \ref{tab:extended_orbit_substructure} and find that they coincide with regions of particularly low average $J_z$ as compared to other stars at similar $|z|$.

The strongest features---e.g. the \texttt{green (C1)} and \texttt{orange (G1)} ridge (see Table \ref{tab:extended_orbit_substructure})---show up in all the panels of Figure \ref{fig:mean_vertical_action}. The low-$J_R$ substructure is only visible out to $|z| \sim 500~\text{pc}$ both in stellar number densities (lower left panels of Figure \ref{fig:mean_vertical_action}) and in the mean $J_z$ (compare the different $J_z$ ranges covered by the colour bars of Panels \ref{fig:mean_vertical_action_S} and \ref{fig:mean_vertical_action_M}; note, however, that measurement uncertainties are larger at high $|z|$ in Panel \ref{fig:mean_vertical_action_L}). This has two implications. First, it appears that the substructure in $(L_z,J_R)$ is indeed physical and not due to, e.g., measurement uncertainties. Measurement uncertainties could explain stars at large $J_R$, but not why the same stars should have at the same time small $J_z$ (see fig. 1 in \citealt{2018MNRAS.481.2970C}). Secondly, the substructure seems to affect stars over a range of $J_z$, but is more prominent when $J_z$ is lower (see also the discussion in Section \ref{sec:nature_action_structure}).

An analogous investigation of the variance of $J_z$ in each $(L_z,\sqrt{J_R})$ bin showed a similar but weaker trend of the action features having also a lower variance, which is not surprising given the high number of stars with low $J_z$ in the substructure.

In general, Figure \ref{fig:mean_vertical_action} shows a trend that at low $L_z$ the average $J_z$ is higher than at high $L_z$. This can be explained by the radial decrease of the vertical velocity dispersion $\sigma_z(R)$ (cf. \citealt{2012ApJ...755..115B}, and the $J_z$-term in Equation \eqref{eq:qdf_guide}). A slight trend of high-$J_R$ stars also having higher $J_z$ in reality, as well as selection effects at the low-$J_R$ edge (see Figure \ref{fig:mock_data}), might also play a role.

\section{Discussion and Conclusion} \label{sec:discussion_conclusion}

\subsection{Summary}

In this work, we have shown that there is a system of orbit substructure in the Galactic disc stars that reaches consistently beyond the Solar neighbourhood out to $\sim 1.5$~kpc. This has been possible for the first time, thanks to \emph{Gaia} DR2/RVS's high-precision measurements of millions of stellar positions and velocities. The distribution of orbital actions $n(J_R,L_z)$ exhibits a wealth of clumps and ridges. At low radial action $J_R$ these map to the known $UVW$ moving groups and their arch-like extensions \citep{2018A&A...616A..11G} in the Solar neighbourhood but are strongly affected by the restricted survey volume. 

In addition, there are also ridge-like features pointing towards high $J_R$ with slopes $\Delta J_R/\Delta L_z \in [0.7,0.9]$, which are not compact $UVW$ clumps in the Solar neighbourhood. We are confident that these $n(J_R,L_z)$ features are physical as (i) we have observed the same features in all distance regimes out to $1/\varpi=1.5~\text{kpc}$, as well as outside and inside, in front and behind of the Sun in the Galactic plane; (ii) the typical measurement uncertainties are smaller for this sample than the size of the observed structures; (iii) these overdensities in $n(J_R,L_z)$ have low mean $J_z$, i.e., stay close to the disc; and (iv) many are related to regions of highly asymmetric radial motion (in-out imbalance). These features reflect the large-scale orbital substructure of the Galactic disc, which appears to be intricate throughout the observed volume.

We now resummarize some of the empirical findings, along with a cautious and preliminary interpretation.

\subsection{The orbit structure beyond the Solar neighbourhood} \label{sec:discussion_orbit_structure}

Figures \ref{fig:sol_neighbourhood} and \ref{fig:action_arrow_velocity} have illustrated that beyond the Solar neighbourhood $(\sim200~\text{pc})$ velocity space is less suitable to reveal orbit substructure as compared to action space, for plausible reasons: disc orbits allow three coordinates to describe the position on (or the phase of) an orbit (e.g. the angles) and three conjugate momenta to label the orbit (e.g. the actions). In the Solar neighbourhood, all stars have basically the same position $\mathbf{x}$, so the conjugate momenta, the velocities $\mathbf{v}$, are good proxies for orbit labels. In large spatial regions, velocities at widely different positions cannot be compared to each other in a meaningful way. There are different approaches to circumvent this. (i) The velocity distributions are independently investigated and modelled in different, small spatial bins (see e.g., \citealt{2010ApJ...725.1676B,2012MNRAS.426L...1A,2013MNRAS.430.3276M,2018A&A...616A..11G}). (ii) Resorting to action-angle space, where orbital phases and labels can be globally investigated and modelled (see e.g., \citealt{2010MNRAS.409..145S,2011MNRAS.418.1565M}). 

One example of the former is fig. 24 in \citet{2018A&A...616A..11G}, which shows the \emph{Gaia} DR2/RVS velocity space in several small spatial bins beyond the Solar neighbourhood. They note that the location of the overdensities in velocity space varies with location within the Galaxy. Here---in particular in Figure \ref{fig:action_distribution_inside_outside}---we have demonstrated that this is a consequence of looking at the \emph{same} orbit structure at different $(R,\phi)$ and therefore $(v_R,v_T)$. Figure \ref{fig:action_distribution_inside_outside}, however, also showed that the strength of the features appeared to vary slightly with $(R,\phi)$, suggesting (together with Figure \ref{fig:voronoi}) that the angle coordinates still contain additional information. Ultimately, it appears that an orbit description, e.g. through actions and angles, is the more sensible approach beyond the small volumes that Hipparcos data had dictated.

\subsection{Asymmetric radial motions}

In Section \ref{sec:asymmetric_vR}, we showed the dramatic asymmetries in stellar number counts in $v_R\approx -U_\text{LSR}$, as a function of $(J_R,L_z)$.  We refer the reader to the illuminating fig. 7 in \citet{2011MNRAS.418.1565M}, which shows the relation between the orbital phase angles $(\theta_R,\theta_\phi)$ and the $(U_\text{LSR},V_\text{LSR})$ velocities. A population phase-mixed in the radial phases $\theta_R$ should therefore have a symmetric distribution in $v_R$. The distribution of azimuthal phase $\theta_\phi$ especially is, however---as \citet{2011MNRAS.418.1565M} shows---strongly affected by selection effects.
This radial motion imbalance therefore implies that many orbits in the Galactic disc (7-9~kpc) are either not phase-mixed along the orbit, or on resonant orbits. 

The next step in the study of orbit substructure in \emph{Gaia} DR2/RVS would be to explicitly calculate and investigate the angles. In a follow-up work of this study, \citet{2018arXiv181003325S} show the angle distribution of the \emph{Gaia} DR2/RVS stars within $200~\text{pc}$. They note that for this restricted volume any substructure in the actions and/or velocities is also bound to show up in the angles. A previous suggestion by \citet{2010MNRAS.409..145S} that for resonant stars the angle coordinates are expected to follow simple relations could not be confirmed with the \emph{Gaia} DR2 data. Further progress needs to be made to understand the effect of resonances on the distribution of orbital angles. We emphasize, however, that we expect the asymmetry in the number density of stars in $v_R$ or $\theta_R$ along each orbit to be highly informative about the nature of the substructure.

In any case, knowing and correcting for the survey selection function will be crucial when interpreting velocity, action, and/or angle distributions. 

\subsection{The nature of the action substructure} \label{sec:nature_action_structure}

We have been able to show the extended system of orbit substructure within $1.5~\text{kpc}$ from the Sun revealed by the \emph{Gaia} DR2/RVS data. This leaves us to explain the ``nature'' of the various ridges that we observed in action space.

We illustrated how the action substructure is related to the known moving groups and the newly discovered arches in velocity space within $200~\text{pc}$ \citep{2018A&A...616A..11G}. In particular, the Hyades, Hercules, and Sirius are related to, or part of (i) high-$J_R$ ridges and (ii) regions of highly asymmetric motions. We deduce that their appearance in $(U,V)$ is the local, selection-effect-affected manifestation of the extended orbital action structure. The Pleiades and the Coma Berenices group (i) contain open star clusters, but have no high-$J_R$ counterpart, (ii) appear phase-mixed in $v_R$, and (iii) are located along the selection effect low-$J_R$ edge of the action distribution in the Solar neighbourhood, and (iv) show in the case of Coma Berenices signs of incomplete vertical phase mixing \citep{2018RNAAS...2b..32M}. This suggests that they do not belong to the extended in-plane orbit structure and have different origins.

What causes the action ridges and corresponding velocity arches beyond the moving groups? \citet{2018A&A...616A..11G} noted that arch-like structures similar to those in the local \emph{Gaia} DR2 velocity data were found to be caused by resonances of the bar in simulations by \citet{2000AJ....119..800D} and \citet{2001A&A...373..511F}. In the action/angle-based perturbation studies by \citet{2017MNRAS.465.1443M,2017MNRAS.471.4314M}, the bar caused an arch-like shearing in velocity space. Studies that investigated the effect of resonances of bars and spiral arms in action space \citep{2002MNRAS.336..785S,2010MNRAS.409..145S,2012ApJ...751...44S,2011MNRAS.418.1565M,2015MNRAS.449.1967F,2015ApJ...806..117F,2018arXiv181003325S} found that resonances cause narrow ridges at lines approximately following constant $\Delta J_R/\Delta L_z$ in action space. Using various approximations of the Fokker-Planck equation, \citet{2015MNRAS.449.1967F,2015ApJ...806..117F} demonstrated that these ridges coincide with directions of rapid orbit diffusion (for tightly wound spirals). Stars located near the $L_z$ corresponding to the radius of the resonance, and which had originally low-eccentricity orbits, move on orbits that become progressively more eccentric (i.e., larger $J_R$) and/or change their guiding-centre radius (i.e., changing $L_z$) due to the resonance, causing the steep ridges in $n(L_z,J_R)$. The exact direction of the ridge depends on the underlying resonance 
\citep[e.g. ILR, OLR or corotation,][]{2002MNRAS.336..785S,2015MNRAS.449.1982F,2018arXiv181003325S}, but the behaviour is quite complex. The strength of the ridge depends (i) on the strength of the diffusion coefficient, which in turn depends on the position in orbit space, and (ii) on the velocity dispersion of the underlying tracers, with kinematically cold populations in the Galactic disc being affected the most. This could be an explanation why we see the action substructure most prominently at low $J_z$, i.e., in the in-plane orbits (see Section \ref{sec:mean_Jz} and also \citealt{2010MNRAS.409..145S}). 

Recently, \citet{2018MNRAS.474.2706B} has shown, based on perturbation theory, that stars in a barred potential are expected to oscillate around the location of the resonances in the $(L_z,J_R)$ plane estimated from an axisymmetric potential. The observed axisymmetric action features will therefore always depend on the current orbital phase of the stars and the location of the survey volume within the Galaxy and with respect to the bar.

Inspired by the \emph{Gaia} DR2 data and the rich action substructure presented in this work, \citet{2018arXiv181003325S} investigate the qualitative effect of different spiral arm models on the action-angle distribution of the Solar neighbourhood. A similar study, Hunt et al. (in preparation), focuses on the effect of the bar in combination with spiral arms. Both studies find that bars and spiral arms could be indeed the cause of the observed features. But the resulting structures are various and model-dependent, and a quantitative explanation of the observed substructure will remain a challenge for the future.

Many of the above-mentioned studies investigate resonance effects on originally smooth, phase-mixed disc distributions. Last but not least, it is therefore worth mentioning that it is quite likely that the Milky Way disc had actually never ever reached a state of complete phase-mixing. Signatures of the ongoing phase-mixing after specific star formation events might still be visible in the data, and structures caused by transient perturbations like satellite fly-bys \citep{2018Natur.561..360A} or stochastic spiral arms \citep{2004MNRAS.350..627D,2018MNRAS.481.3794H} might be currently in the process of mixing back in.

\subsection{Caveats}

One caveat when investigating actions is the uncertainty in our knowledge about the overall shape of the Milky Way's gravitational potential. We have performed some tests assuming different (wrong) shapes for the potential, the action substructure was in each case still clearly discernible. This is partly due to $L_z = R \times v_T$ depending on the potential only weakly via $v_T \sim V_\text{LSR} + v_\text{circ}(R_{\odot})$. The observed slopes $\Delta J_R/\Delta L_z$ and the exact locations of the overdensities in the $(L_z,J_R)$ plane do rely on the assumed potential. However, due to the noisy, blurred-out appearance of the overdensities and the limited power of our ``by-eye" fitting method, we estimate the uncertainty of the slopes listed for the \texttt{MWPotential2014} in Table \ref{tab:extended_orbit_substructure} conservatively to 20\%. For example, varying the scale parameters of the Miyamoto-Nagai disc and NFW halo by 50\% appeared to change the slopes by amounts of less than these 20\%. The $J_R$ location was shifted by up to $15\%$. In any case, if the slopes are informative about the specific mechanism causing each feature---as the work by, e.g., \citet{2002MNRAS.336..785S} and \citet{2018arXiv181003325S} might suggest---progress in our understanding of either the potential or the underlying mechanisms should therefore help to learn more about the other.

In the long term, we do indeed expect that \emph{Gaia} data will help to improve measurements of the Galaxy's gravitational potential, allowing more accurate action estimation (e.g., \citealt{2016ApJ...830...97T,2018arXiv180411348W,2018arXiv180501408P}). 

The exact location of the features in action space also depends on the choice of the Solar motion with respect to the Local Standard of Rest. In this work, we use $(U_\odot,V_\odot,W_\odot) = (11.1, 12.24, 7.25)~\text{km s}^{-1}$ by \citet{2010MNRAS.403.1829S}. We have also performed the tests in this work using the measurement by \citet{2015ApJ...809..145T}, $(U_\odot,V_\odot,W_\odot) = (9.58, 10.52, 7.01)~\text{km s}^{-1}$, from LAMOST data. The action features shift up to $\Delta L_z/L_{z,0} \sim 0.02$ along the lower $J_R$ edge of the distribution. Qualitatively, the shape of the structures does not change, however.

Also, while in an axisymmetric potential the actions are well defined and integrals of motions, the Galaxy is clearly not axisymmetric. Any action calculation per se will therefore only be an action \emph{estimation}. We emphasize that we have treated the actions in this work simply as an informative coordinate system to characterize the orbits that the stars currently are on, \emph{under the assumption} of an axisymmetric gravitational potential model for the Milky Way (\texttt{MWPotential2014} by \citealt{2015ApJS..216...29B}). Secular evolution and orbital diffusion will need to be evoked in the future to explain the observed orbital structure. Recent work by \citet{2018MNRAS.474.2706B} and \citet{2018arXiv181003325S} on (perturbed) actions for resonant orbits suggests that axisymmetric actions can still be informative. We will explore this further in future studies.

\subsection{Concluding remarks}

It is clear that the rich structure in action space contains a wealth of information about non-axisymmetric perturbations and resonances in the Galactic disc. The fact that Figure \ref{fig:sol_neighbourhood} reveals (at least) seven separate ridges at different orientations in the $(L_z,J_R)$ plane suggests that more than one (resonance) mechanism might be at work. Action space alone will most likely not be sufficient to determine which kinds of resonances are at work. Angle space and chemical abundances, as well as a good knowledge of the selection function, might help to disentangle the effects in the future. The modelling of this substructure is however highly complex and beyond the scope of this paper. 

In the past, action estimations have largely been dominated by measurement uncertainties \citep{2018MNRAS.481.2970C} and substructure in action space was forgiving when using axisymmetric models to describe the Galaxy \citep{2017ApJ...839...61T}. \emph{Gaia} DR2 might finally mark the beginning of an era where the limitations of our models to capture the complexity of the data will be the limitations to our knowledge about the Milky Way.

\section*{Acknowledgments}

The authors thank the anonymous referee for helpful comments, as well as Jerry Sellwood, Paul McMillan, Simon White, Amina Helmi, James Binney, and David Hogg for useful feedback on early versions of this work. J.C.  acknowledges support from the SFB 881 program (A3) and the International Max Planck Research School for Astronomy and Cosmic Physics at Heidelberg University (IMPRS-HD).  H.W.R. received support from the European Research Council under the European Union's Seventh Framework Programme (FP7) ERC Grant Agreement n. [321035]. This work has made use of data from the European Space Agency (ESA) mission {\it Gaia} (\url{https://www.cosmos.esa.int/Gaia}), processed by the {\it Gaia} Data Processing and Analysis Consortium (DPAC, \url{https://www.cosmos.esa.int/web/Gaia/dpac/consortium}). Funding for the DPAC has been provided by national institutions, in particular the institutions participating in the {\it Gaia} Multilateral Agreement.

\appendix

\begin{figure*}
    \centering
    \includegraphics[width=0.9\textwidth]{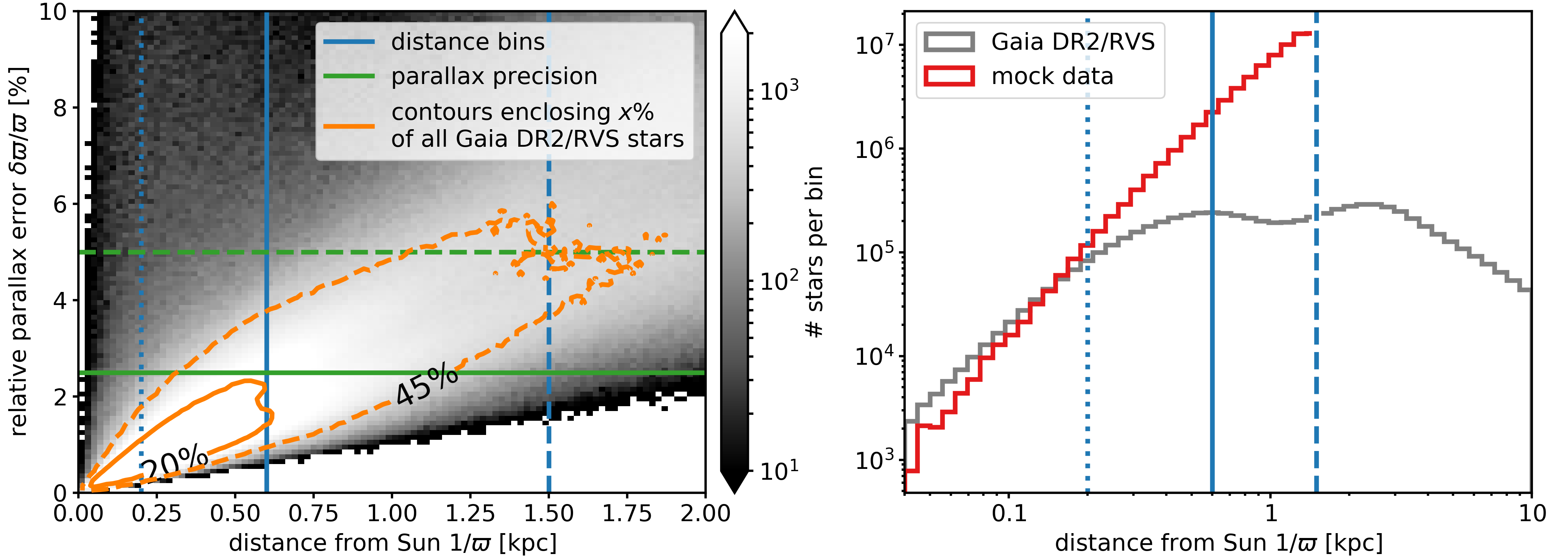}
    \caption{Motivating the restriction of the \emph{Gaia} DR2/RVS sample to stars within $1/\varpi < 1.5~\text{kpc}$ from the Sun in this work. The blue vertical lines mark the distance bins $[200, 600, 1500]~\text{pc}$ we use in Figures \ref{fig:sol_neighbourhood}, \ref{fig:error_ellipses}, \ref{fig:action_arrow_velocity}, and \ref{fig:voronoi}. The left-hand panel shows the star distribution in relative parallax error versus the distance. A restriction to $1.5~\text{kpc}$ ($600~\text{pc}$) still retains 45\% (20\%) of all stars that have distance errors smaller than 5\% (2.5\%). The right-hand panel compares the number of stars as a function of distance from the Sun with a (scaled, complete) disc mock data set (see Section \ref{fig:mock_data}). This should give a rough feeling for the completeness of the star sample. Within $200~\text{pc}$ the slopes are similar. Within $600~\text{pc}$ and $1.5~\text{kpc}$ the RVS sample contains already less stars than the model prediction. The strong drop in numbers outside of $2~\text{kpc}$ is due to the magnitude limit of the RVS \citep{2018arXiv180409372K}.}
    \label{fig:distance_bins}
\end{figure*}


\bibliographystyle{mnras}
\bibliography{bibliography} 

\begin{thebibliography}{}
\makeatletter
\relax
\def\mn@urlcharsother{\let\do\@makeother \do\$\do\&\do\#\do\^\do\_\do\%\do\~}
\def\mn@doi{\begingroup\mn@urlcharsother \@ifnextchar [ {\mn@doi@}
  {\mn@doi@[]}}
\def\mn@doi@[#1]#2{\def\@tempa{#1}\ifx\@tempa\@empty \href
  {http://dx.doi.org/#2} {doi:#2}\else \href {http://dx.doi.org/#2} {#1}\fi
  \endgroup}
\def\mn@eprint#1#2{\mn@eprint@#1:#2::\@nil}
\def\mn@eprint@arXiv#1{\href {http://arxiv.org/abs/#1} {{\tt arXiv:#1}}}
\def\mn@eprint@dblp#1{\href {http://dblp.uni-trier.de/rec/bibtex/#1.xml}
  {dblp:#1}}
\def\mn@eprint@#1:#2:#3:#4\@nil{\def\@tempa {#1}\def\@tempb {#2}\def\@tempc
  {#3}\ifx \@tempc \@empty \let \@tempc \@tempb \let \@tempb \@tempa \fi \ifx
  \@tempb \@empty \def\@tempb {arXiv}\fi \@ifundefined
  {mn@eprint@\@tempb}{\@tempb:\@tempc}{\expandafter \expandafter \csname
  mn@eprint@\@tempb\endcsname \expandafter{\@tempc}}}

\bibitem[\protect\citeauthoryear{{Antoja}, {Figueras}, {Fern{\'a}ndez}  \&
  {Torra}}{{Antoja} et~al.}{2008}]{2008A&A...490..135A}
{Antoja} T.,  {Figueras} F.,  {Fern{\'a}ndez} D.,   {Torra} J.,  2008, \mn@doi
  [\aap] {10.1051/0004-6361:200809519}, \href
  {http://adsabs.harvard.edu/abs/2008A%26A...490..135A} {490, 135}

\bibitem[\protect\citeauthoryear{{Antoja}, {Valenzuela}, {Pichardo}, {Moreno},
  {Figueras}  \& {Fern{\'a}ndez}}{{Antoja} et~al.}{2009}]{2009ApJ...700L..78A}
{Antoja} T.,  {Valenzuela} O.,  {Pichardo} B.,  {Moreno} E.,  {Figueras} F.,
  {Fern{\'a}ndez} D.,  2009, \mn@doi [\apjl] {10.1088/0004-637X/700/2/L78},
  \href {http://adsabs.harvard.edu/abs/2009ApJ...700L..78A} {700, L78}

\bibitem[\protect\citeauthoryear{{Antoja}, {Figueras}, {Torra}, {Valenzuela}
  \& {Pichardo}}{{Antoja} et~al.}{2010}]{2010LNEA....4...13B}
{Antoja} T.,  {Figueras} F.,  {Torra} J.,  {Valenzuela} O.,   {Pichardo} B.,
  2010, Lecture Notes and Essays in Astrophysics, \href
  {http://adsabs.harvard.edu/abs/2010LNEA....4...13B} {4, 13}

\bibitem[\protect\citeauthoryear{{Antoja}, {Figueras}, {Romero-G{\'o}mez},
  {Pichardo}, {Valenzuela}  \& {Moreno}}{{Antoja}
  et~al.}{2011}]{2011MNRAS.418.1423A}
{Antoja} T.,  {Figueras} F.,  {Romero-G{\'o}mez} M.,  {Pichardo} B.,
  {Valenzuela} O.,   {Moreno} E.,  2011, \mn@doi [\mnras]
  {10.1111/j.1365-2966.2011.19190.x}, \href
  {http://adsabs.harvard.edu/abs/2011MNRAS.418.1423A} {418, 1423}

\bibitem[\protect\citeauthoryear{{Antoja} et~al.,}{{Antoja}
  et~al.}{2012}]{2012MNRAS.426L...1A}
{Antoja} T.,  et~al., 2012, \mn@doi [\mnras]
  {10.1111/j.1745-3933.2012.01310.x}, \href
  {http://adsabs.harvard.edu/abs/2012MNRAS.426L...1A} {426, L1}

\bibitem[\protect\citeauthoryear{{Antoja} et~al.,}{{Antoja}
  et~al.}{2014}]{2014A&A...563A..60A}
{Antoja} T.,  et~al., 2014, \mn@doi [\aap] {10.1051/0004-6361/201322623}, \href
  {http://adsabs.harvard.edu/abs/2014A%26A...563A..60A} {563, A60}

\bibitem[\protect\citeauthoryear{{Antoja} et~al.,}{{Antoja}
  et~al.}{2015}]{2015ApJ...800L..32A}
{Antoja} T.,  et~al., 2015, \mn@doi [\apjl] {10.1088/2041-8205/800/2/L32},
  \href {http://adsabs.harvard.edu/abs/2015ApJ...800L..32A} {800, L32}

\bibitem[\protect\citeauthoryear{{Antoja} et~al.,}{{Antoja}
  et~al.}{2018}]{2018Natur.561..360A}
{Antoja} T.,  et~al., 2018, \mn@doi [\nat] {10.1038/s41586-018-0510-7}, \href
  {http://adsabs.harvard.edu/abs/2018Natur.561..360A} {561, 360}

\bibitem[\protect\citeauthoryear{{Arifyanto} \& {Fuchs}}{{Arifyanto} \&
  {Fuchs}}{2006}]{2006A&A...449..533A}
{Arifyanto} M.~I.,  {Fuchs} B.,  2006, \mn@doi [\aap]
  {10.1051/0004-6361:20054355}, \href
  {http://adsabs.harvard.edu/abs/2006A%26A...449..533A} {449, 533}

\bibitem[\protect\citeauthoryear{{Bailer-Jones}, {Rybizki}, {Fouesneau},
  {Mantelet}  \& {Andrae}}{{Bailer-Jones} et~al.}{2018}]{2018AJ....156...58B}
{Bailer-Jones} C.~A.~L.,  {Rybizki} J.,  {Fouesneau} M.,  {Mantelet} G.,
  {Andrae} R.,  2018, \mn@doi [\aj] {10.3847/1538-3881/aacb21}, \href
  {http://adsabs.harvard.edu/abs/2018AJ....156...58B} {156, 58}

\bibitem[\protect\citeauthoryear{{Bensby}, {Oey}, {Feltzing}  \&
  {Gustafsson}}{{Bensby} et~al.}{2007}]{2007ApJ...655L..89B}
{Bensby} T.,  {Oey} M.~S.,  {Feltzing} S.,   {Gustafsson} B.,  2007, \mn@doi
  [\apjl] {10.1086/512014}, \href
  {http://adsabs.harvard.edu/abs/2007ApJ...655L..89B} {655, L89}

\bibitem[\protect\citeauthoryear{{Bensby}, {Feltzing}  \& {Oey}}{{Bensby}
  et~al.}{2014}]{2014A&A...562A..71B}
{Bensby} T.,  {Feltzing} S.,   {Oey} M.~S.,  2014, \mn@doi [\aap]
  {10.1051/0004-6361/201322631}, \href
  {http://adsabs.harvard.edu/abs/2014A%26A...562A..71B} {562, A71}

\bibitem[\protect\citeauthoryear{{Binney}}{{Binney}}{2012}]{2012MNRAS.426.1324B}
{Binney} J.,  2012, \mn@doi [\mnras] {10.1111/j.1365-2966.2012.21757.x}, \href
  {http://adsabs.harvard.edu/abs/2012MNRAS.426.1324B} {426, 1324}

\bibitem[\protect\citeauthoryear{{Binney}}{{Binney}}{2018}]{2018MNRAS.474.2706B}
{Binney} J.,  2018, \mn@doi [\mnras] {10.1093/mnras/stx2835}, \href
  {http://adsabs.harvard.edu/abs/2018MNRAS.474.2706B} {474, 2706}

\bibitem[\protect\citeauthoryear{{Binney} \& {Lacey}}{{Binney} \&
  {Lacey}}{1988}]{1988MNRAS.230..597B}
{Binney} J.,  {Lacey} C.,  1988, \mn@doi [\mnras] {10.1093/mnras/230.4.597},
  \href {https://ui.adsabs.harvard.edu/#abs/1988MNRAS.230..597B} {230, 597}

\bibitem[\protect\citeauthoryear{{Binney} \& {McMillan}}{{Binney} \&
  {McMillan}}{2011}]{2011MNRAS.413.1889B}
{Binney} J.,  {McMillan} P.,  2011, \mn@doi [\mnras]
  {10.1111/j.1365-2966.2011.18268.x}, \href
  {http://adsabs.harvard.edu/abs/2011MNRAS.413.1889B} {413, 1889}

\bibitem[\protect\citeauthoryear{{Binney} \& {Piffl}}{{Binney} \&
  {Piffl}}{2015}]{2015MNRAS.454.3653B}
{Binney} J.,  {Piffl} T.,  2015, \mn@doi [\mnras] {10.1093/mnras/stv2225},
  \href {http://adsabs.harvard.edu/abs/2015MNRAS.454.3653B} {454, 3653}

\bibitem[\protect\citeauthoryear{{Binney} \& {Sch{\"o}nrich}}{{Binney} \&
  {Sch{\"o}nrich}}{2018}]{2018MNRAS.481.1501B}
{Binney} J.,  {Sch{\"o}nrich} R.,  2018, \mn@doi [\mnras]
  {10.1093/mnras/sty2378}, \href
  {http://adsabs.harvard.edu/abs/2018MNRAS.481.1501B} {481, 1501}

\bibitem[\protect\citeauthoryear{{Binney} \& {Tremaine}}{{Binney} \&
  {Tremaine}}{2008}]{2008gady.book.....B}
{Binney} J.,  {Tremaine} S.,  2008, {Galactic Dynamics: Second Edition}.
Princeton University Press

\bibitem[\protect\citeauthoryear{{Bland-Hawthorn} et~al.,}{{Bland-Hawthorn}
  et~al.}{2019}]{2019MNRAS.tmp..222B}
{Bland-Hawthorn} J.,  et~al., 2019, \mn@doi [\mnras] {10.1093/mnras/stz217},
  \href {http://adsabs.harvard.edu/abs/2019MNRAS.tmp..222B} {}

\bibitem[\protect\citeauthoryear{{Bovy}}{{Bovy}}{2010}]{2010ApJ...725.1676B}
{Bovy} J.,  2010, \mn@doi [\apj] {10.1088/0004-637X/725/2/1676}, \href
  {http://adsabs.harvard.edu/abs/2010ApJ...725.1676B} {725, 1676}

\bibitem[\protect\citeauthoryear{{Bovy}}{{Bovy}}{2015}]{2015ApJS..216...29B}
{Bovy} J.,  2015, \mn@doi [\apjs] {10.1088/0067-0049/216/2/29}, \href
  {http://adsabs.harvard.edu/abs/2015ApJS..216...29B} {216, 29}

\bibitem[\protect\citeauthoryear{{Bovy} \& {Hogg}}{{Bovy} \&
  {Hogg}}{2010}]{2010ApJ...717..617B}
{Bovy} J.,  {Hogg} D.~W.,  2010, \mn@doi [\apj] {10.1088/0004-637X/717/2/617},
  \href {http://adsabs.harvard.edu/abs/2010ApJ...717..617B} {717, 617}

\bibitem[\protect\citeauthoryear{{Bovy} \& {Rix}}{{Bovy} \&
  {Rix}}{2013}]{2013ApJ...779..115B}
{Bovy} J.,  {Rix} H.-W.,  2013, \mn@doi [\apj] {10.1088/0004-637X/779/2/115},
  \href {http://adsabs.harvard.edu/abs/2013ApJ...779..115B} {779, 115}

\bibitem[\protect\citeauthoryear{{Bovy}, {Hogg}  \& {Roweis}}{{Bovy}
  et~al.}{2009}]{2009ApJ...700.1794B}
{Bovy} J.,  {Hogg} D.~W.,   {Roweis} S.~T.,  2009, \mn@doi [\apj]
  {10.1088/0004-637X/700/2/1794}, \href
  {http://adsabs.harvard.edu/abs/2009ApJ...700.1794B} {700, 1794}

\bibitem[\protect\citeauthoryear{{Bovy}, {Rix}, {Hogg}, {Beers}, {Lee}  \&
  {Zhang}}{{Bovy} et~al.}{2012a}]{2012ApJ...755..115B}
{Bovy} J.,  {Rix} H.-W.,  {Hogg} D.~W.,  {Beers} T.~C.,  {Lee} Y.~S.,   {Zhang}
  L.,  2012a, \mn@doi [\apj] {10.1088/0004-637X/755/2/115}, \href
  {http://adsabs.harvard.edu/abs/2012ApJ...755..115B} {755, 115}

\bibitem[\protect\citeauthoryear{{Bovy} et~al.,}{{Bovy}
  et~al.}{2012b}]{2012ApJ...759..131B}
{Bovy} J.,  et~al., 2012b, \mn@doi [\apj] {10.1088/0004-637X/759/2/131}, \href
  {http://adsabs.harvard.edu/abs/2012ApJ...759..131B} {759, 131}

\bibitem[\protect\citeauthoryear{{Chakrabarty}}{{Chakrabarty}}{2007}]{2007A&A...467..145C}
{Chakrabarty} D.,  2007, \mn@doi [\aap] {10.1051/0004-6361:20066677}, \href
  {http://adsabs.harvard.edu/abs/2007A%26A...467..145C} {467, 145}

\bibitem[\protect\citeauthoryear{{Coronado}, {Rix}  \& {Trick}}{{Coronado}
  et~al.}{2018}]{2018MNRAS.481.2970C}
{Coronado} J.,  {Rix} H.-W.,   {Trick} W.~H.,  2018, \mn@doi [\mnras]
  {10.1093/mnras/sty2468}, \href
  {http://adsabs.harvard.edu/abs/2018MNRAS.481.2970C} {481, 2970}

\bibitem[\protect\citeauthoryear{{Das} \& {Binney}}{{Das} \&
  {Binney}}{2016}]{2016MNRAS.460.1725D}
{Das} P.,  {Binney} J.,  2016, \mn@doi [\mnras] {10.1093/mnras/stw744}, \href
  {http://adsabs.harvard.edu/abs/2016MNRAS.460.1725D} {460, 1725}

\bibitem[\protect\citeauthoryear{{Das}, {Williams}  \& {Binney}}{{Das}
  et~al.}{2016}]{2016MNRAS.463.3169D}
{Das} P.,  {Williams} A.,   {Binney} J.,  2016, \mn@doi [\mnras]
  {10.1093/mnras/stw2167}, \href
  {http://adsabs.harvard.edu/abs/2016MNRAS.463.3169D} {463, 3169}

\bibitem[\protect\citeauthoryear{{De Simone}, {Wu}  \& {Tremaine}}{{De Simone}
  et~al.}{2004}]{2004MNRAS.350..627D}
{De Simone} R.,  {Wu} X.,   {Tremaine} S.,  2004, \mn@doi [\mnras]
  {10.1111/j.1365-2966.2004.07675.x}, \href
  {http://adsabs.harvard.edu/abs/2004MNRAS.350..627D} {350, 627}

\bibitem[\protect\citeauthoryear{{Dehnen}}{{Dehnen}}{1998}]{1998AJ....115.2384D}
{Dehnen} W.,  1998, \mn@doi [\aj] {10.1086/300364}, \href
  {http://adsabs.harvard.edu/abs/1998AJ....115.2384D} {115, 2384}

\bibitem[\protect\citeauthoryear{{Dehnen}}{{Dehnen}}{2000}]{2000AJ....119..800D}
{Dehnen} W.,  2000, \mn@doi [\aj] {10.1086/301226}, \href
  {http://adsabs.harvard.edu/abs/2000AJ....119..800D} {119, 800}

\bibitem[\protect\citeauthoryear{{Eggen}}{{Eggen}}{1958}]{1958MNRAS.118...65E}
{Eggen} O.~J.,  1958, \mn@doi [\mnras] {10.1093/mnras/118.1.65}, \href
  {http://adsabs.harvard.edu/abs/1958MNRAS.118...65E} {118, 65}

\bibitem[\protect\citeauthoryear{{Eggen}}{{Eggen}}{1996}]{1996AJ....112.1595E}
{Eggen} O.~J.,  1996, \mn@doi [\aj] {10.1086/118126}, \href
  {http://adsabs.harvard.edu/abs/1996AJ....112.1595E} {112, 1595}

\bibitem[\protect\citeauthoryear{{Famaey}, {Jorissen}, {Luri}, {Mayor}, {Udry},
  {Dejonghe}  \& {Turon}}{{Famaey} et~al.}{2005}]{2005A&A...430..165F}
{Famaey} B.,  {Jorissen} A.,  {Luri} X.,  {Mayor} M.,  {Udry} S.,  {Dejonghe}
  H.,   {Turon} C.,  2005, \mn@doi [\aap] {10.1051/0004-6361:20041272}, \href
  {http://adsabs.harvard.edu/abs/2005A%26A...430..165F} {430, 165}

\bibitem[\protect\citeauthoryear{{Famaey}, {Pont}, {Luri}, {Udry}, {Mayor}  \&
  {Jorissen}}{{Famaey} et~al.}{2007}]{2007A&A...461..957F}
{Famaey} B.,  {Pont} F.,  {Luri} X.,  {Udry} S.,  {Mayor} M.,   {Jorissen} A.,
  2007, \mn@doi [\aap] {10.1051/0004-6361:20065706}, \href
  {http://adsabs.harvard.edu/abs/2007A%26A...461..957F} {461, 957}

\bibitem[\protect\citeauthoryear{{Famaey}, {Siebert}  \& {Jorissen}}{{Famaey}
  et~al.}{2008}]{2008A&A...483..453F}
{Famaey} B.,  {Siebert} A.,   {Jorissen} A.,  2008, \mn@doi [\aap]
  {10.1051/0004-6361:20078979}, \href
  {http://adsabs.harvard.edu/abs/2008A%26A...483..453F} {483, 453}

\bibitem[\protect\citeauthoryear{{Fouvry} \& {Pichon}}{{Fouvry} \&
  {Pichon}}{2015}]{2015MNRAS.449.1982F}
{Fouvry} J.-B.,  {Pichon} C.,  2015, \mn@doi [\mnras] {10.1093/mnras/stv360},
  \href {https://ui.adsabs.harvard.edu/#abs/2015MNRAS.449.1982F} {449, 1982}

\bibitem[\protect\citeauthoryear{{Fouvry}, {Pichon}  \& {Prunet}}{{Fouvry}
  et~al.}{2015a}]{2015MNRAS.449.1967F}
{Fouvry} J.-B.,  {Pichon} C.,   {Prunet} S.,  2015a, \mn@doi [\mnras]
  {10.1093/mnras/stv359}, \href
  {http://adsabs.harvard.edu/abs/2015MNRAS.449.1967F} {449, 1967}

\bibitem[\protect\citeauthoryear{{Fouvry}, {Binney}  \& {Pichon}}{{Fouvry}
  et~al.}{2015b}]{2015ApJ...806..117F}
{Fouvry} J.-B.,  {Binney} J.,   {Pichon} C.,  2015b, \mn@doi [\apj]
  {10.1088/0004-637X/806/1/117}, \href
  {http://adsabs.harvard.edu/abs/2015ApJ...806..117F} {806, 117}

\bibitem[\protect\citeauthoryear{{Fux}}{{Fux}}{2001}]{2001A&A...373..511F}
{Fux} R.,  2001, \mn@doi [\aap] {10.1051/0004-6361:20010561}, \href
  {http://adsabs.harvard.edu/abs/2001A%26A...373..511F} {373, 511}

\bibitem[\protect\citeauthoryear{{Gaia Collaboration} et~al.,}{{Gaia
  Collaboration} et~al.}{2016}]{2016A&A...595A...1G}
{Gaia Collaboration} et~al., 2016, \mn@doi [\aap]
  {10.1051/0004-6361/201629272}, \href
  {http://adsabs.harvard.edu/abs/2016A%26A...595A...1G} {595, A1}

\bibitem[\protect\citeauthoryear{{Gaia Collaboration} et~al.,}{{Gaia
  Collaboration} et~al.}{2018a}]{2018A&A...616A...1G}
{Gaia Collaboration} et~al., 2018a, \mn@doi [\aap]
  {10.1051/0004-6361/201833051}, \href
  {http://adsabs.harvard.edu/abs/2018A%26A...616A...1G} {616, A1}

\bibitem[\protect\citeauthoryear{{Gaia Collaboration} et~al.,}{{Gaia
  Collaboration} et~al.}{2018b}]{2018A&A...616A..11G}
{Gaia Collaboration} et~al., 2018b, \mn@doi [\aap]
  {10.1051/0004-6361/201832865}, \href
  {http://adsabs.harvard.edu/abs/2018A%26A...616A..11G} {616, A11}

\bibitem[\protect\citeauthoryear{{Hattori}, {Gouda}, {Tagawa}, {Sakai}, {Yano},
  {Baba}  \& {Kumamoto}}{{Hattori} et~al.}{2019}]{2019MNRAS.tmp..263H}
{Hattori} K.,  {Gouda} N.,  {Tagawa} H.,  {Sakai} N.,  {Yano} T.,  {Baba} J.,
  {Kumamoto} J.,  2019, \mn@doi [\mnras] {10.1093/mnras/stz266}, \href
  {http://adsabs.harvard.edu/abs/2019MNRAS.tmp..263H} {}

\bibitem[\protect\citeauthoryear{{Helmi}, {White}, {de Zeeuw}  \&
  {Zhao}}{{Helmi} et~al.}{1999}]{1999Natur.402...53H}
{Helmi} A.,  {White} S.~D.~M.,  {de Zeeuw} P.~T.,   {Zhao} H.,  1999, \mn@doi
  [\nat] {10.1038/46980}, \href
  {http://adsabs.harvard.edu/abs/1999Natur.402...53H} {402, 53}

\bibitem[\protect\citeauthoryear{{H{\o}g} et~al.,}{{H{\o}g}
  et~al.}{2000}]{2000A&A...355L..27H}
{H{\o}g} E.,  et~al., 2000, \aap, \href
  {http://adsabs.harvard.edu/abs/2000A%26A...355L..27H} {355, L27}

\bibitem[\protect\citeauthoryear{{Holmberg}, {Nordstr{\"o}m}  \&
  {Andersen}}{{Holmberg} et~al.}{2009}]{2009A&A...501..941H}
{Holmberg} J.,  {Nordstr{\"o}m} B.,   {Andersen} J.,  2009, \mn@doi [\aap]
  {10.1051/0004-6361/200811191}, \href
  {http://adsabs.harvard.edu/abs/2009A%26A...501..941H} {501, 941}

\bibitem[\protect\citeauthoryear{{Hunt} \& {Bovy}}{{Hunt} \&
  {Bovy}}{2018}]{2018MNRAS.477.3945H}
{Hunt} J.~A.~S.,  {Bovy} J.,  2018, \mn@doi [\mnras] {10.1093/mnras/sty921},
  \href {http://adsabs.harvard.edu/abs/2018MNRAS.477.3945H} {477, 3945}

\bibitem[\protect\citeauthoryear{{Hunt}, {Hong}, {Bovy}, {Kawata}  \&
  {Grand}}{{Hunt} et~al.}{2018}]{2018MNRAS.481.3794H}
{Hunt} J.~A.~S.,  {Hong} J.,  {Bovy} J.,  {Kawata} D.,   {Grand} R.~J.~J.,
  2018, \mn@doi [\mnras] {10.1093/mnras/sty2532}, \href
  {http://adsabs.harvard.edu/abs/2018MNRAS.481.3794H} {481, 3794}

\bibitem[\protect\citeauthoryear{{Juri{\'c}} et~al.,}{{Juri{\'c}}
  et~al.}{2008}]{2008ApJ...673..864J}
{Juri{\'c}} M.,  et~al., 2008, \mn@doi [\apj] {10.1086/523619}, \href
  {http://adsabs.harvard.edu/abs/2008ApJ...673..864J} {673, 864}

\bibitem[\protect\citeauthoryear{{Katz} et~al.,}{{Katz}
  et~al.}{2018}]{2018arXiv180409372K}
{Katz} D.,  et~al., 2018, preprint, \href
  {http://adsabs.harvard.edu/abs/2018arXiv180409372K} {} (\mn@eprint {arXiv}
  {1804.09372})

\bibitem[\protect\citeauthoryear{{Kawata}, {Baba}, {Ciuc{\v a}}, {Cropper},
  {Grand}, {Hunt}  \& {Seabroke}}{{Kawata} et~al.}{2018}]{2018MNRAS.479L.108K}
{Kawata} D.,  {Baba} J.,  {Ciuc{\v a}} I.,  {Cropper} M.,  {Grand} R.~J.~J.,
  {Hunt} J.~A.~S.,   {Seabroke} G.,  2018, \mn@doi [\mnras]
  {10.1093/mnrasl/sly107}, \href
  {http://adsabs.harvard.edu/abs/2018MNRAS.479L.108K} {479, L108}

\bibitem[\protect\citeauthoryear{{Klement}, {Fuchs}  \& {Rix}}{{Klement}
  et~al.}{2008}]{2008ApJ...685..261K}
{Klement} R.,  {Fuchs} B.,   {Rix} H.-W.,  2008, \mn@doi [\apj]
  {10.1086/590139}, \href {http://adsabs.harvard.edu/abs/2008ApJ...685..261K}
  {685, 261}

\bibitem[\protect\citeauthoryear{{Koppelman}, {Virginiflosia}, {Posti},
  {Veljanoski}  \& {Helmi}}{{Koppelman} et~al.}{2018}]{2018arXiv180407530K}
{Koppelman} H.~H.,  {Virginiflosia} T.,  {Posti} L.,  {Veljanoski} J.,
  {Helmi} A.,  2018, preprint, \href
  {http://adsabs.harvard.edu/abs/2018arXiv180407530K} {} (\mn@eprint {arXiv}
  {1804.07530})

\bibitem[\protect\citeauthoryear{{Lindegren} et~al.,}{{Lindegren}
  et~al.}{2018}]{2018A&A...616A...2L}
{Lindegren} L.,  et~al., 2018, \mn@doi [\aap] {10.1051/0004-6361/201832727},
  \href {http://adsabs.harvard.edu/abs/2018A%26A...616A...2L} {616, A2}

\bibitem[\protect\citeauthoryear{{Luri} et~al.,}{{Luri}
  et~al.}{2018}]{2018A&A...616A...9L}
{Luri} X.,  et~al., 2018, \mn@doi [\aap] {10.1051/0004-6361/201832964}, \href
  {http://adsabs.harvard.edu/abs/2018A%26A...616A...9L} {616, A9}

\bibitem[\protect\citeauthoryear{{McMillan}}{{McMillan}}{2011}]{2011MNRAS.418.1565M}
{McMillan} P.~J.,  2011, \mn@doi [\mnras] {10.1111/j.1365-2966.2011.19520.x},
  \href {http://adsabs.harvard.edu/abs/2011MNRAS.418.1565M} {418, 1565}

\bibitem[\protect\citeauthoryear{{McMillan}}{{McMillan}}{2013}]{2013MNRAS.430.3276M}
{McMillan} P.~J.,  2013, \mn@doi [\mnras] {10.1093/mnras/stt129}, \href
  {http://adsabs.harvard.edu/abs/2013MNRAS.430.3276M} {430, 3276}

\bibitem[\protect\citeauthoryear{{Minchev}, {Quillen}, {Williams}, {Freeman},
  {Nordhaus}, {Siebert}  \& {Bienaym{\'e}}}{{Minchev}
  et~al.}{2009}]{2009MNRAS.396L..56M}
{Minchev} I.,  {Quillen} A.~C.,  {Williams} M.,  {Freeman} K.~C.,  {Nordhaus}
  J.,  {Siebert} A.,   {Bienaym{\'e}} O.,  2009, \mn@doi [\mnras]
  {10.1111/j.1745-3933.2009.00661.x}, \href
  {http://adsabs.harvard.edu/abs/2009MNRAS.396L..56M} {396, L56}

\bibitem[\protect\citeauthoryear{{Minchev}, {Boily}, {Siebert}  \&
  {Bienayme}}{{Minchev} et~al.}{2010}]{2010MNRAS.407.2122M}
{Minchev} I.,  {Boily} C.,  {Siebert} A.,   {Bienayme} O.,  2010, \mn@doi
  [\mnras] {10.1111/j.1365-2966.2010.17060.x}, \href
  {http://adsabs.harvard.edu/abs/2010MNRAS.407.2122M} {407, 2122}

\bibitem[\protect\citeauthoryear{{Monari}, {Famaey}  \& {Siebert}}{{Monari}
  et~al.}{2016}]{2016MNRAS.457.2569M}
{Monari} G.,  {Famaey} B.,   {Siebert} A.,  2016, \mn@doi [\mnras]
  {10.1093/mnras/stw171}, \href
  {http://adsabs.harvard.edu/abs/2016MNRAS.457.2569M} {457, 2569}

\bibitem[\protect\citeauthoryear{{Monari}, {Famaey}, {Siebert}, {Duchateau},
  {Lorscheider}  \& {Bienaym{\'e}}}{{Monari}
  et~al.}{2017a}]{2017MNRAS.465.1443M}
{Monari} G.,  {Famaey} B.,  {Siebert} A.,  {Duchateau} A.,  {Lorscheider} T.,
  {Bienaym{\'e}} O.,  2017a, \mn@doi [\mnras] {10.1093/mnras/stw2807}, \href
  {http://adsabs.harvard.edu/abs/2017MNRAS.465.1443M} {465, 1443}

\bibitem[\protect\citeauthoryear{{Monari}, {Kawata}, {Hunt}  \&
  {Famaey}}{{Monari} et~al.}{2017b}]{2017MNRAS.466L.113M}
{Monari} G.,  {Kawata} D.,  {Hunt} J.~A.~S.,   {Famaey} B.,  2017b, \mn@doi
  [\mnras] {10.1093/mnrasl/slw238}, \href
  {http://adsabs.harvard.edu/abs/2017MNRAS.466L.113M} {466, L113}

\bibitem[\protect\citeauthoryear{{Monari}, {Famaey}, {Fouvry}  \&
  {Binney}}{{Monari} et~al.}{2017c}]{2017MNRAS.471.4314M}
{Monari} G.,  {Famaey} B.,  {Fouvry} J.-B.,   {Binney} J.,  2017c, \mn@doi
  [\mnras] {10.1093/mnras/stx1825}, \href
  {http://adsabs.harvard.edu/abs/2017MNRAS.471.4314M} {471, 4314}

\bibitem[\protect\citeauthoryear{{Monari} et~al.,}{{Monari}
  et~al.}{2018}]{2018RNAAS...2b..32M}
{Monari} G.,  et~al., 2018, \mn@doi [Research Notes of the American
  Astronomical Society] {10.3847/2515-5172/aac38e}, \href
  {http://adsabs.harvard.edu/abs/2018RNAAS...2b..32M} {2, 32}

\bibitem[\protect\citeauthoryear{{Moreno}, {Pichardo}  \& {Schuster}}{{Moreno}
  et~al.}{2015}]{2015MNRAS.451..705M}
{Moreno} E.,  {Pichardo} B.,   {Schuster} W.~J.,  2015, \mn@doi [\mnras]
  {10.1093/mnras/stv962}, \href
  {http://adsabs.harvard.edu/abs/2015MNRAS.451..705M} {451, 705}

\bibitem[\protect\citeauthoryear{{Navarro}, {Helmi}  \& {Freeman}}{{Navarro}
  et~al.}{2004}]{2004ApJ...601L..43N}
{Navarro} J.~F.,  {Helmi} A.,   {Freeman} K.~C.,  2004, \mn@doi [\apjl]
  {10.1086/381751}, \href {http://adsabs.harvard.edu/abs/2004ApJ...601L..43N}
  {601, L43}

\bibitem[\protect\citeauthoryear{{P{\'e}rez-Villegas}, {Portail}, {Wegg}  \&
  {Gerhard}}{{P{\'e}rez-Villegas} et~al.}{2017}]{2017ApJ...840L...2P}
{P{\'e}rez-Villegas} A.,  {Portail} M.,  {Wegg} C.,   {Gerhard} O.,  2017,
  \mn@doi [\apjl] {10.3847/2041-8213/aa6c26}, \href
  {http://adsabs.harvard.edu/abs/2017ApJ...840L...2P} {840, L2}

\bibitem[\protect\citeauthoryear{{Piffl} et~al.,}{{Piffl}
  et~al.}{2014}]{2014MNRAS.445.3133P}
{Piffl} T.,  et~al., 2014, \mn@doi [\mnras] {10.1093/mnras/stu1948}, \href
  {http://adsabs.harvard.edu/abs/2014MNRAS.445.3133P} {445, 3133}

\bibitem[\protect\citeauthoryear{{Pomp{\'e}ia} et~al.,}{{Pomp{\'e}ia}
  et~al.}{2011}]{2011MNRAS.415.1138P}
{Pomp{\'e}ia} L.,  et~al., 2011, \mn@doi [\mnras]
  {10.1111/j.1365-2966.2011.18685.x}, \href
  {http://adsabs.harvard.edu/abs/2011MNRAS.415.1138P} {415, 1138}

\bibitem[\protect\citeauthoryear{{Posti} \& {Helmi}}{{Posti} \&
  {Helmi}}{2018}]{2018arXiv180501408P}
{Posti} L.,  {Helmi} A.,  2018, preprint, \href
  {http://adsabs.harvard.edu/abs/2018arXiv180501408P} {} (\mn@eprint {arXiv}
  {1805.01408})

\bibitem[\protect\citeauthoryear{{Posti}, {Binney}, {Nipoti}  \&
  {Ciotti}}{{Posti} et~al.}{2015}]{2015MNRAS.447.3060P}
{Posti} L.,  {Binney} J.,  {Nipoti} C.,   {Ciotti} L.,  2015, \mn@doi [\mnras]
  {10.1093/mnras/stu2608}, \href
  {http://adsabs.harvard.edu/abs/2015MNRAS.447.3060P} {447, 3060}

\bibitem[\protect\citeauthoryear{{Quillen}}{{Quillen}}{2003}]{2003AJ....125..785Q}
{Quillen} A.~C.,  2003, \mn@doi [\aj] {10.1086/345725}, \href
  {http://adsabs.harvard.edu/abs/2003AJ....125..785Q} {125, 785}

\bibitem[\protect\citeauthoryear{{Quillen} \& {Minchev}}{{Quillen} \&
  {Minchev}}{2005}]{2005AJ....130..576Q}
{Quillen} A.~C.,  {Minchev} I.,  2005, \mn@doi [\aj] {10.1086/430885}, \href
  {http://adsabs.harvard.edu/abs/2005AJ....130..576Q} {130, 576}

\bibitem[\protect\citeauthoryear{{Quillen}, {Dougherty}, {Bagley}, {Minchev}
  \& {Comparetta}}{{Quillen} et~al.}{2011}]{2011MNRAS.417..762Q}
{Quillen} A.~C.,  {Dougherty} J.,  {Bagley} M.~B.,  {Minchev} I.,
  {Comparetta} J.,  2011, \mn@doi [\mnras] {10.1111/j.1365-2966.2011.19349.x},
  \href {http://adsabs.harvard.edu/abs/2011MNRAS.417..762Q} {417, 762}

\bibitem[\protect\citeauthoryear{{Sanders} \& {Binney}}{{Sanders} \&
  {Binney}}{2015}]{2015MNRAS.449.3479S}
{Sanders} J.~L.,  {Binney} J.,  2015, \mn@doi [\mnras] {10.1093/mnras/stv578},
  \href {http://adsabs.harvard.edu/abs/2015MNRAS.449.3479S} {449, 3479}

\bibitem[\protect\citeauthoryear{{Sanders} \& {Binney}}{{Sanders} \&
  {Binney}}{2016}]{2016MNRAS.457.2107S}
{Sanders} J.~L.,  {Binney} J.,  2016, \mn@doi [\mnras] {10.1093/mnras/stw106},
  \href {http://adsabs.harvard.edu/abs/2016MNRAS.457.2107S} {457, 2107}

\bibitem[\protect\citeauthoryear{{Sch{\"o}nrich} \& {Dehnen}}{{Sch{\"o}nrich}
  \& {Dehnen}}{2018}]{2018MNRAS.478.3809S}
{Sch{\"o}nrich} R.,  {Dehnen} W.,  2018, \mn@doi [\mnras]
  {10.1093/mnras/sty1256}, \href
  {http://adsabs.harvard.edu/abs/2018MNRAS.478.3809S} {478, 3809}

\bibitem[\protect\citeauthoryear{{Sch{\"o}nrich}, {Binney}  \&
  {Dehnen}}{{Sch{\"o}nrich} et~al.}{2010}]{2010MNRAS.403.1829S}
{Sch{\"o}nrich} R.,  {Binney} J.,   {Dehnen} W.,  2010, \mn@doi [\mnras]
  {10.1111/j.1365-2966.2010.16253.x}, \href
  {http://adsabs.harvard.edu/abs/2010MNRAS.403.1829S} {403, 1829}

\bibitem[\protect\citeauthoryear{{Sellwood}}{{Sellwood}}{2010}]{2010MNRAS.409..145S}
{Sellwood} J.~A.,  2010, \mn@doi [\mnras] {10.1111/j.1365-2966.2010.17305.x},
  \href {http://adsabs.harvard.edu/abs/2010MNRAS.409..145S} {409, 145}

\bibitem[\protect\citeauthoryear{{Sellwood}}{{Sellwood}}{2012}]{2012ApJ...751...44S}
{Sellwood} J.~A.,  2012, \mn@doi [\apj] {10.1088/0004-637X/751/1/44}, \href
  {http://adsabs.harvard.edu/abs/2012ApJ...751...44S} {751, 44}

\bibitem[\protect\citeauthoryear{{Sellwood} \& {Binney}}{{Sellwood} \&
  {Binney}}{2002}]{2002MNRAS.336..785S}
{Sellwood} J.~A.,  {Binney} J.~J.,  2002, \mn@doi [\mnras]
  {10.1046/j.1365-8711.2002.05806.x}, \href
  {http://adsabs.harvard.edu/abs/2002MNRAS.336..785S} {336, 785}

\bibitem[\protect\citeauthoryear{{Sellwood}, {Trick}, {Carlberg}, {Coronado}
  \& {Rix}}{{Sellwood} et~al.}{2018}]{2018arXiv181003325S}
{Sellwood} J.~A.,  {Trick} W.,  {Carlberg} R.,  {Coronado} J.,   {Rix} H.-W.,
  2018, preprint, \href {http://adsabs.harvard.edu/abs/2018arXiv181003325S} {}
  (\mn@eprint {arXiv} {1810.03325})

\bibitem[\protect\citeauthoryear{{Solway}, {Sellwood}  \&
  {Sch{\"o}nrich}}{{Solway} et~al.}{2012}]{2012MNRAS.422.1363S}
{Solway} M.,  {Sellwood} J.~A.,   {Sch{\"o}nrich} R.,  2012, \mn@doi [\mnras]
  {10.1111/j.1365-2966.2012.20712.x}, \href
  {http://adsabs.harvard.edu/abs/2012MNRAS.422.1363S} {422, 1363}

\bibitem[\protect\citeauthoryear{{Tian} et~al.,}{{Tian}
  et~al.}{2015}]{2015ApJ...809..145T}
{Tian} H.-J.,  et~al., 2015, \mn@doi [\apj] {10.1088/0004-637X/809/2/145},
  \href {http://adsabs.harvard.edu/abs/2015ApJ...809..145T} {809, 145}

\bibitem[\protect\citeauthoryear{{Trick}}{{Trick}}{2017}]{phdthesis}
{Trick} W.~H.,  2017, dissertation, MPI for Astronomy/Heidelberg University,
  \mn@doi{10.11588/heidok.00023767}

\bibitem[\protect\citeauthoryear{{Trick}, {Bovy}  \& {Rix}}{{Trick}
  et~al.}{2016}]{2016ApJ...830...97T}
{Trick} W.~H.,  {Bovy} J.,   {Rix} H.-W.,  2016, \mn@doi [\apj]
  {10.3847/0004-637X/830/2/97}, \href
  {http://adsabs.harvard.edu/abs/2016ApJ...830...97T} {830, 97}

\bibitem[\protect\citeauthoryear{{Trick}, {Bovy}, {D'Onghia}  \& {Rix}}{{Trick}
  et~al.}{2017}]{2017ApJ...839...61T}
{Trick} W.~H.,  {Bovy} J.,  {D'Onghia} E.,   {Rix} H.-W.,  2017, \mn@doi [\apj]
  {10.3847/1538-4357/aa67db}, \href
  {http://adsabs.harvard.edu/abs/2017ApJ...839...61T} {839, 61}

\bibitem[\protect\citeauthoryear{{Vera-Ciro} \& {D'Onghia}}{{Vera-Ciro} \&
  {D'Onghia}}{2016}]{2016ApJ...824...39V}
{Vera-Ciro} C.,  {D'Onghia} E.,  2016, \mn@doi [\apj]
  {10.3847/0004-637X/824/1/39}, \href
  {http://adsabs.harvard.edu/abs/2016ApJ...824...39V} {824, 39}

\bibitem[\protect\citeauthoryear{{Watkins}, {van der Marel}, {Sohn}  \&
  {Evans}}{{Watkins} et~al.}{2018}]{2018arXiv180411348W}
{Watkins} L.~L.,  {van der Marel} R.~P.,  {Sohn} S.~T.,   {Evans} N.~W.,  2018,
  preprint, \href {http://adsabs.harvard.edu/abs/2018arXiv180411348W} {}
  (\mn@eprint {arXiv} {1804.11348})

\bibitem[\protect\citeauthoryear{{Williams}, {Freeman}, {Helmi}  \& {RAVE
  Collaboration}}{{Williams} et~al.}{2009}]{2009IAUS..254..139W}
{Williams} M.~E.~K.,  {Freeman} K.~C.,  {Helmi} A.,   {RAVE Collaboration}
  2009, in {Andersen} J.,  {Nordstr{\"o}ara} {m} B.,   {Bland-Hawthorn} J.,
  eds,  IAU Symposium Vol. 254, The Galaxy Disk in Cosmological Context. pp
  139--144 (\mn@eprint {arXiv} {0810.2669}), \mn@doi{10.1017/S1743921308027518}

\bibitem[\protect\citeauthoryear{{Zhao}, {Zhao}, {Chen}, {Oswalt}, {Tan}  \&
  {Zhang}}{{Zhao} et~al.}{2014}]{2014ApJ...787...31Z}
{Zhao} J.~K.,  {Zhao} G.,  {Chen} Y.~Q.,  {Oswalt} T.~D.,  {Tan} K.~F.,
  {Zhang} Y.,  2014, \mn@doi [\apj] {10.1088/0004-637X/787/1/31}, \href
  {http://adsabs.harvard.edu/abs/2014ApJ...787...31Z} {787, 31}

\makeatother
\end{thebibliography}


\section{Data quality}\label{sec:data_quality}

Figure \ref{fig:distance_bins} motivates our choice to restrict the investigation in this work to stars with $1/\varpi < 1.5~\text{kpc}$.

First, this is a good compromise between retaining a significant fraction of stars of the DR2/RVS sample for our investigation ($\gtrsim45\%$) and assuring precise parallax measurements (mostly less than 5\% relative error; left-hand panel of Figure \ref{fig:distance_bins}). Inside of $1/\varpi=600~\text{pc}$, most stars even have less than 2.5\% parallax uncertainties.

Secondly, this distance cut reduces problems due to the data incompleteness resulting from the magnitude limit of the RVS spectrograph $G\sim12~\text{mag}$ (see e.g., fig. 6 in \citealt{2018arXiv180409372K}). Sun-like stars with $G\sim5~\text{mag}$ should be therefore contained in the sample out to $d\sim250~\text{pc}$ and red clump giants with $G\sim0.5~\text{mag}$ out to $d\sim2~\text{kpc}$. And indeed, by comparing a (very idealized and complete) mock data prediction for the expected stellar numbers out to $1.5~\text{kpc}$ (see Section \ref{sec:mock_data}) with the data in the right-hand panel of Figure \ref{fig:distance_bins}, we see that the slopes within $200~\text{pc}$ are very similar, suggesting high completeness in the Solar neighbourhood. The decrease in stellar numbers outside of $1/\varpi\sim2~\text{kpc}$ is due to the aforementioned magnitude limit. The variations in the number density slope between $1/\varpi \approx 0.6$ and $2.5~\text{kpc}$ cannot be reproduced by the mock data and are most likely caused by the RVS sample's incompleteness as a function of parallax, and related to the excess of stars at $R<R_\odot$.

We have experimented with different quality cuts. (i) Figure \ref{fig:distance_bins} shows that close stars can also have large parallax uncertainties. This is due to the \emph{Gaia} scanning law and crowding. Cutting our sample down in \texttt{parallax\_over\_error} reduces the data set very unevenly over the plane of the sky, e.g. up to 80\% in the direction of the Galactic centre. (ii) We also tested a quality cut in the error of the 3D space velocity (i.e., the combination of $v_\text{los}$ and the proper motion velocity vector), and (iii) in the \texttt{astrometric\_excess\_noise} to remove binaries from the sample. The latter two produced however insignificant reductions to the data set after the parallax error cut had already been performed.

Most importantly, all (qualitative) results in this work appeared unchanged, even when not performing any quality cuts. We therefore show action space for all \emph{Gaia} DR2/RVS stars within $1/\varpi < 1.5~\text{kpc}$ to not impose any selection biases other than inherent to the \emph{Gaia} data anyway.

\section{Short review on moving groups in the Solar neighbourhood} \label{sec:reviewing_moving_groups}

The $(U,V)$ plane of the Solar neighbourhood has long revealed overdensities of stars that appear to move on common trajectories \citep{1998AJ....115.2384D}. Among them are the Sirius, Coma Berenices, the Hyades and Pleiades, and the Hercules stream (see Figure \ref{fig:UV_arrow_action}, left-hand panel). Notwithstanding the star clusters of the same name, these groups do not correspond to overdensities in space.

Several origins have been proposed for these moving groups. (a) They could be open clusters, born together, but no longer gravitationally bound, and now dispersing (see series of papers by Eggen, starting with \citet{1958MNRAS.118...65E}, and also references therein). (b) They could be accreted and disrupted subhalos, as predicted by hierarchical galaxy formation theory \mbox{\citep{1999Natur.402...53H}}. The Arcturus stream at $V\sim -100~\text{km s}^{-1}$ and spreading a wide range in $U$ \citep{2009IAUS..254..139W}, for example, appears to have an extragalactic origin \citep{2004ApJ...601L..43N,2014ApJ...787...31Z}. (c) They could be the local manifestation of dynamic resonance effects caused by non-axisymmetric perturbations in the Galactic disc, like the bar \citep{2000AJ....119..800D,2010MNRAS.407.2122M}, or spiral waves \citep{2004MNRAS.350..627D,2005AJ....130..576Q,2011MNRAS.418.1423A}, or an overlap of both \citep{2003AJ....125..785Q,2007A&A...467..145C,2009ApJ...700L..78A,2011MNRAS.417..762Q,2015MNRAS.451..705M}. (d) Another dynamical origin could be perturbations of the disc due to subhalo mergers \citep{2009MNRAS.396L..56M,2018RNAAS...2b..32M}.

Some of the moving groups appear to have a wide range of stellar ages \citep{2005A&A...430..165F,2008A&A...490..135A} and average metallicities that stand out from the stellar background distribution\footnote{The average metallicity of the Hyades \citep{2011MNRAS.415.1138P,2010ApJ...717..617B} and Hercules \citep{2010ApJ...717..617B,2014A&A...562A..71B,2019MNRAS.tmp..263H} is slightly higher, and those of Sirius lower \citep{2010ApJ...717..617B} than those of the local thin-disc population. This suggests that these stars originate from the inner or outer disc, respectively, and have been scattered into the Solar neighbourhood by dynamic processes.} \citep{2007ApJ...655L..89B,2014A&A...562A..71B,2010ApJ...717..617B,2011MNRAS.415.1138P}---both consistent with a dynamic rather than a common birth origin. In particular, the Hyades group might be caused by an inner Lindblad resonance of spiral arms \citep{2005AJ....130..576Q,2010MNRAS.409..145S,2011MNRAS.418.1565M,2013MNRAS.430.3276M}. The Hercules stream could be due to the 2:1 outer Lindblad resonance (OLR) of a short, fast bar \citep{2000AJ....119..800D,2014A&A...563A..60A,2017MNRAS.466L.113M}, or created by a long, slow bar's corotation resonance \citep{2017ApJ...840L...2P} or 4:1 OLR \citep{2018MNRAS.477.3945H}, or in combination with transient spiral arms \citep{2018MNRAS.481.3794H}. Coma Berenices is only present at negative Galactic latitudes, which suggests that it could have been caused by the passage of a dwarf galaxy \citep{2018RNAAS...2b..32M}.

In addition to the $(U,V)$ plane, the moving groups could also be identified as overdensities in orbital quantities, like energy, angular momentum, eccentricity, etc. (e.g., \citealt{2006A&A...449..533A,2008ApJ...685..261K}).

Before \emph{Gaia} DR2, quantification of the extent of moving groups in the $(U,V)$ plane was more difficult, due to the lesser quality and quantity of existing data sets (e.g. \citealt{2009ApJ...700.1794B}). \citet{2018A&A...616A..11G} have just recently identified arches with nearly constant $V$ and spanning a wide range of $U$ in \emph{Gaia} DR2 data within $d=200~\text{pc}$, which coincide with known moving groups.


\bsp	
\label{lastpage}
\end{document}